\documentclass[pdflatex,sn-mathphys,referee]{sn-jnl}
\usepackage{siunitx}
\DeclareSIUnit\angstrom{\text {Å}}

\usepackage{physics}

\jyear{2022}
\unnumbered

\newcommand{\BSO}{\vec{B}_\mathrm{SO}}
\newcommand{\VLO}{V_\mathrm{LO}}
\newcommand{\VLD}{V_\mathrm{LD}}
\newcommand{\VLI}{V_\mathrm{LI}}
\newcommand{\VRI}{V_\mathrm{RI}}
\newcommand{\VRD}{V_\mathrm{RD}}
\newcommand{\VRO}{V_\mathrm{RO}}
\newcommand{\VL}{V_\mathrm{L}}
\newcommand{\VR}{V_\mathrm{R}}
\newcommand{\IL}{I_\mathrm{L}}
\newcommand{\IR}{I_\mathrm{R}}
\newcommand{\Icor}{I_\mathrm{corr}}
\newcommand{\Icar}{I_\mathrm{CAR}}
\newcommand{\Iect}{I_\mathrm{ECT}}
\newcommand{\GRR}{G_\mathrm{RR}}
\newcommand{\GLL}{G_\mathrm{LL}}
\newcommand{\GRL}{G_\mathrm{RL}}
\newcommand{\GLR}{G_\mathrm{LR}}
\newcommand{\VPG}{V_\mathrm{PG}}
\newcommand{\muL}{\mu_\mathrm{LD}}
\newcommand{\muR}{\mu_\mathrm{RD}}

\newcommand{\DU}{\ket{\mathrm{D}\!\uparrow}}
\newcommand{\DD}{\ket{\mathrm{D}\!\downarrow}}
\newcommand{\EABS}{E_\mathrm{ABS}}
\newcommand{\EZ}{E_\mathrm{Z}}

\newcommand{\nn}{\nonumber \\} 

\newcommand{\dg}{^{\dagger}}
\newcommand{\hc}{\rm{H.c.}}

\begin{document}

\title[Controlled crossed Andreev reflection and elastic co-tunneling]{Controlled crossed Andreev reflection and elastic co-tunneling mediated by Andreev~bound~states}

\author[1]{\fnm{Alberto}~\sur{Bordin}}
\equalcont{These authors contributed equally to this work.}

\author[1]{\fnm{Guanzhong}~\sur{Wang}}
\equalcont{These authors contributed equally to this work.}

\author[1]{\fnm{Chun-Xiao}~\sur{Liu}}
\equalcont{These authors contributed equally to this work.}

\author[1]{\fnm{Sebastiaan~L.~D.}~\sur{ten~Haaf}}

\author[1]{\fnm{Grzegorz~P.}~\sur{Mazur}}

\author[1]{\fnm{Nick}~\sur{van~Loo}}

\author[1]{\fnm{Di}~\sur{Xu}}

\author[1]{\fnm{David}~\sur{van~Driel}}

\author[1]{\fnm{Francesco}~\sur{Zatelli}}


\author[2]{\fnm{Sasa}~\sur{Gazibegovic}}

\author[2]{\fnm{Ghada}~\sur{Badawy}}

\author[2]{\fnm{Erik~P.~A.~M.}~\sur{Bakkers}}

\author[1]{\fnm{Michael}~\sur{Wimmer}}

\author*[1]{\fnm{Leo~P.}~\sur{Kouwenhoven}}\email{L.P.Kouwenhoven@tudelft.nl}

\author*[1]{\fnm{Tom}~\sur{Dvir}}\email{tom.dvir@gmail.com}

\affil[1]{\orgdiv{QuTech and Kavli Institute of NanoScience}, \orgname{Delft University of Technology}, \postcode{2600 GA} \orgaddress{\city{Delft}, \country{The Netherlands}}}

\affil[2]{\orgdiv{Department of Applied Physics}, \orgname{Eindhoven University of Technology}, \postcode{5600 MB} \orgaddress{\city{Eindhoven}, \country{The Netherlands}}}

\abstract{
A short superconducting segment can couple attached quantum dots via elastic co-tunneling (ECT) and crossed Andreev reflection (CAR).
Such coupled quantum dots can host Majorana bound states provided that the ratio between CAR and ECT can be controlled.
Metallic superconductors have so far been shown to mediate such tunneling phenomena, albeit with limited tunability.  
Here we show that Andreev bound states formed in semiconductor-superconductor heterostructures can mediate CAR and ECT over mesoscopic length scales. 
Andreev bound states possess both an electron and a hole component, giving rise to an intricate interference phenomenon that allows us to tune the ratio between CAR and ECT deterministically. 
We further show that the combination of intrinsic spin-orbit coupling in InSb nanowires and an applied magnetic field provides another efficient knob to tune the ratio between ECT and CAR and optimize the amount of coupling between neighboring quantum dots.
}

\maketitle

\begin{figure}[h]
    \centering
    \includegraphics[width = \textwidth]{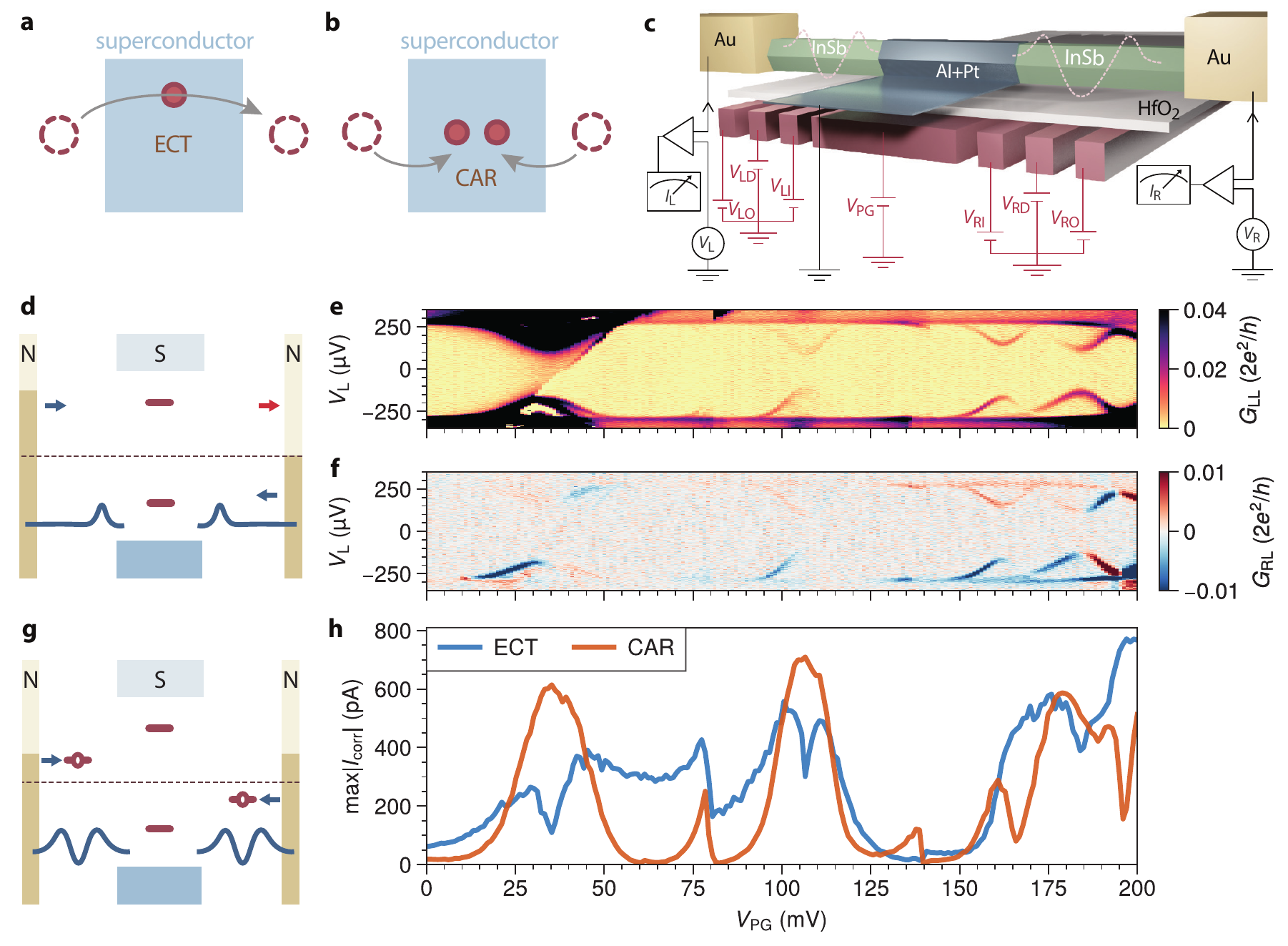}
    \caption{\textbf{Correlation between ABS and CAR/ECT processes.} 
    \textbf{a.} Illustration of the ECT process: a single electron tunnels between two QDs. 
    \textbf{b.} Illustration of CAR: two electrons from two QDs enter the superconducting segment simultaneously to form a Cooper pair.  
    \textbf{c.} Schematic illustration of our devices and experimental setup. An InSb nanowire (green) is coated by a thin Al shell (blue, Al+Pt for device A), on top of seven finger gates (red). Two Cr/Au leads (yellow) are attached to both sides of the wire.
    \textbf{d.} Spectroscopy configuration: applying low voltages to $\VLI$ and $\VRI$ creates a single tunneling barrier on each side of the device, enabling local and nonlocal tunneling spectroscopy of the superconducting segment. Yellow bars depict voltage bias in normal (N) contacts while blue rectangles represent the superconductor (S). Blue curves sketch the desired voltage profile defined with the gates; voltage barriers are not to scale.
    \textbf{e.} $\GLL$ as a function of $\VL$ and $\VPG$ when setting the gates in the tunneling spectroscopy configuration. 
    \textbf{f.} $\GRL$ as a function of $\VL$ and $\VPG$ in the same settings of panel (e). $\GLL$ and $\GRL$ are calculated by taking the numerical derivative after applying a Savitzky-Golay filter of window length 11 and polynomial order 1 to the measured $\IL$ and $\IR$ currents, respectively.
    \textbf{g.} Configuration with QDs: applying low voltages on $\VLO$,$\VLI$ and  $\VRI$,$\VRO$ forms a QD on the left and right side of the superconducting segment. 
    \textbf{h.} CAR- and ECT-induced currents as a function of $\VPG$ measured using the $N \leftrightarrow N+1$ transition in both QDs. The values of $V_\mathrm{LI}$ and $V_\mathrm{RI}$ were kept constant during measurements of panels (e-f) and (h).}
    \label{fig: Correlation}
\end{figure}

The Kitaev chain is a prime example of condensed-matter toy models exhibiting a topological superconducting phase~\cite{Kitaev.2001}.
Practical proposals to construct an artificial Kitaev chain require a set of quantum dots (QDs) separated by narrow superconducting segments~\cite{Sau.2012, Leijnse.2012, Fulga.2013}. 
Such QDs interact via two mechanisms: elastic co-tunneling (ECT) and crossed Andreev reflection (CAR). 
ECT occurs when a single electron tunnels between the two QDs via the superconductor (see schematic in Fig.~\ref{fig: Correlation}a). 
In CAR, electrons from two separate QDs tunnel into the superconductor forming a Cooper pair; or in its reversed process, a Cooper pair is split into two electrons, tunneling to different QDs (see schematic in Fig.~\ref{fig: Correlation}b)~\cite{beckmann.2004, Russo2005, Recher.2001}. 
The balance between CAR- and ECT-induced couplings is crucial for observing the sought-after Majorana zero modes at the boundaries of the Kitaev chain~\cite{Leijnse.2012}.

Andreev bound states (ABSs) in hybrid semiconductor-superconductor heterostructures can mediate CAR and ECT between two neighboring QDs~\cite{Liu.2022}. 
Such states form when a confined semiconducting level is tunnel-coupled to a superconductor. 
Importantly, an ABS excitation can be smoothly tuned from electron-like to hole-like using electrostatic gating~\cite{Schindele.2014, Danon.2020, Menard.2020}. 
The interplay between the electron and hole components of an ABS is predicted to be a key element in controlling CAR and ECT~\cite{Liu.2022}.
Moreover, an external magnetic field affects the ABS energy via Zeeman splitting~\cite{Lee.2014} and influences CAR and ECT amplitudes thereby.
In the presence of spin-orbit coupling, the amplitudes further develop an anisotropic dependence on the magnetic field direction~\cite{Wang.2022}.

In this work, we report on gate tunability of CAR and ECT in hybrid semiconductor-superconductor heterostructures. 
In particular, both processes are correlated with the presence of ABSs in the hybrid. 
By comparing experimental data and our theoretical model, we further show that the observed CAR and ECT amplitudes, respectively, result from constructive and destructive interference of tunneling paths. 
The interference pattern is linked to the charge of the mediating Andreev bound state and can be controlled via tuning the hybrid's chemical potential. 
Finally, we report on the magnetic field dependence of CAR and ECT. 
We show how the CAR and ECT interference patterns are modified through the interplay of the orientation of the magnetic field, the direction of the spin-orbit coupling, the energy of the ABS and its spin-splitting.

Fig.~\ref{fig: Correlation}c shows a schematic depiction of the reported devices and the measurement circuit. 
An InSb nanowire is deposited on pre-fabricated metallic gates (separated from the nanowire by a thin dielectric layer). 
Using the shadow lithography technique~\cite{Heedt:2021_NC, Borsoi:2021_AFM}, a thin superconducting layer is deposited on top of the middle segment of the nanowire. 
Normal contacts are then fabricated on each side of the device.
Details of the fabrication are described in Methods. 
Scanning electron microscope images of reported devices are shown in Fig.~\ref{fig: SEMs}.  
Transport measurements are carried out by applying DC voltage biases on the left and the right contacts ($\VL,\VR$) and measuring the resulting DC currents on both sides ($\IL, \IR$). 
Local ($\GLL = dI_\mathrm{L}/dV_\mathrm{L}$, $\GRR = dI_\mathrm{R}/dV_\mathrm{R}$) and nonlocal ($\GRL  = dI_\mathrm{R}/dV_\mathrm{L}, \GLR  = dI_\mathrm{L}/dV_\mathrm{R}$) conductances were obtained as numerical derivatives of the DC currents unless otherwise specified.  
All measurements are conducted in a dilution refrigerator with a measured electron temperature of $\sim \SI{50}{mK}$. 


We begin by characterizing the spectrum of the hybrid semiconducting-superconducting segment. To measure tunnel spectroscopy, we form a single tunnel barrier on each side of this segment, as shown schematically in Fig.~\ref{fig: Correlation}d. 
Fig.~\ref{fig: Correlation}e shows that at low values of $\VPG$, the spectrum features a hard superconducting gap. 
Increasing $\VPG$ leads to the formation of discrete ABSs under the superconducting film appearing as electron-hole symmetric sub-gap peaks. 
These peaks also appear in the nonlocal conductance (Fig.~\ref{fig: Correlation}f), indicating that the ABSs extend throughout the hybrid segment~\cite{Rosdahl.2018}. 

Next, to measure currents induced by CAR and ECT ($\Icar$ and $\Iect$), we form a QD on each side of the hybrid segment. 
We do so by lowering the voltage applied to the gates next to normal leads while keeping the voltages applied to the gates next to the hybrid segment fixed (see schematics in Fig.~\ref{fig: Correlation}g). 
We characterize the QDs by measuring the gate-dependent and magnetic-field-dependent transport through them and focus in the remainder of this paper on two charge transitions of each QD: from $N$ to $N+1$ electrons and $N+1$ to $N+2$ electron where $N$ is a small even integer (see Fig.~\ref{ED_dotchar}). 

The CAR- and ECT-induced currents are measured using a method introduced in our previous work~\cite{Wang.2022}. 
We briefly summarize the full descriptions in the Methods here.
To measure CAR-induced currents at a given value of $\VPG$, we apply a fixed bias of $\VL = \VR = \SI{70}{\micro V}$ on both leads and scan $\VLD$ and $\VRD$ in a range of $\sim \SI{1}{mV}$ around the charge degeneracy point of each dot. 
Due to energy conservation, CAR-induced currents arise when $\VLD$ and $\VRD$ fulfill the condition that the chemical potentials of both QDs are equal in magnitude with an opposite sign with respect to the Fermi energy (shown schematically in Fig.~\ref{fig: Correlation}g).
We measure $\IL$ and $\IR$ and calculate the correlated current: $\Icor \equiv \mathrm{sgn} \left( \IL \IR \right) \sqrt{ \abs{ \IL \IR }}$. 
We take $\Icar \equiv \max(\Icor)$ to represent the amount of CAR-induced currents at a given value of $\VPG$. 
To measure the amount of ECT-induced currents, we repeat this procedure when applying anti-symmetric bias on both leads: $\VL=-\VR = \SI{70}{\micro V}$. 
We take $\Iect \equiv -\min(\Icor)$ to represent the amount of ECT-induced currents at a given value of $\VPG$.

Fig.~\ref{fig: Correlation}h shows the dependence of $\Icar$ and $\Iect$ on $\VPG$ when both QDs are tuned to the $N\leftrightarrow N+1$ transition (see Fig.~\ref{fig: zerofield_all} for data involving $N+1\leftrightarrow N+2$ transitions and the discussion of the effect of Pauli spin blockade). 
Both currents respond strongly to changes in $\VPG$, suggesting that they originate from processes that involve the hybrid segment. 
$\Icar$, in particular, reaches peak currents at $\VPG$ values where ABSs in the hybrid segment reach a minimal energy. 
In regions of $\VPG$ far from ABSs, $\Icar$ and $\Iect$ are suppressed.
These observations hold for all devices we measured (see Fig.~\ref{fig: devB} for another example).


\begin{figure}
    \centering
    \includegraphics[width = 0.5\textwidth]{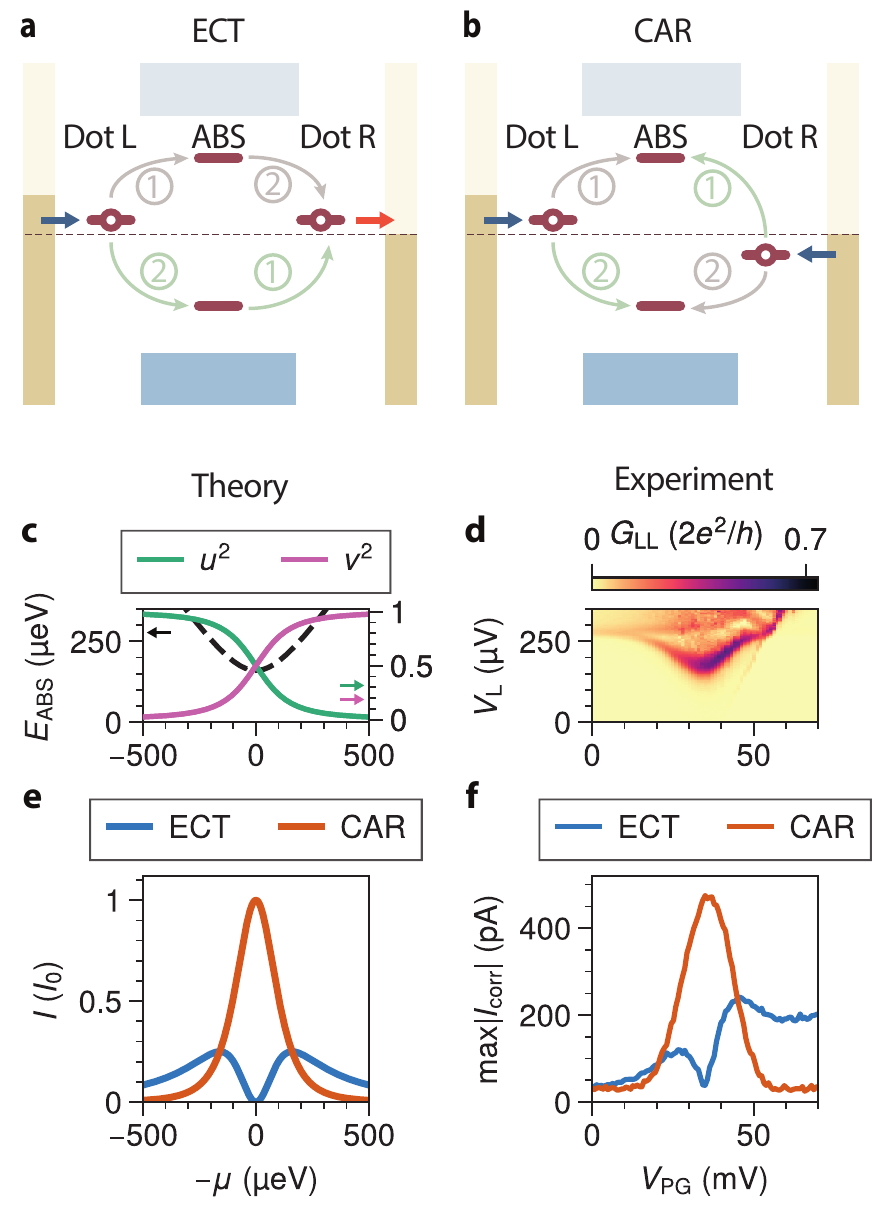}
    \caption{\textbf{Detailed study of CAR and ECT through an ABS.  }
    \textbf{a.} Two possible paths for ECT: an electron hops from the left QD to the center ABS, followed by an escape from the ABS to the right QD (gray), and the processes in the opposite order (green). 
    \textbf{b.} Two possible paths for CAR: an electron from the left QD enters the ABS followed by another electron arriving from the right QD (gray) and the same processes in reversed order (green). 
    \textbf{c.} $E_\mathrm{ABS}$, and $u,v$ as a function of $\mu$ calculated in the atomic limit, where $E_\mathrm{ABS} = \sqrt{\Gamma^2 + \mu^2}$ with $\Gamma = \SI{160}{\micro eV}$ \cite{Bauer.2007}. $\mu$ and $\VPG$ are related via $\mu= -e \alpha (\VPG - V_0)$ where $\alpha$ is the gate lever arm and $V_0 = \SI{35}{mV}$ is an offset. Comparing data to theory, we estimate $\alpha \sim 0.01$.
    \textbf{d.} $\GLL$ as a function of $\VL$ and $\VPG$ showing a single ABS. 
    \textbf{e.} A toy-model calculation of the transmission probability as a function of $\mu$. 
    \textbf{f.} A high-resolution measurement of CAR and ECT amplitudes while tuning $\VPG$. The background noise level is $\sim$\SI{30}{pA} (see Methods).}
    \label{fig: single_ABS}
\end{figure}


To understand the role of ABSs in mediating CAR- and ECT-induced currents, we consider a model with two QDs on each side of a single ABS confined in the central hybrid segment, as shown in Fig.~\ref{fig: single_ABS}a,b. Considering only one orbital state in each QD, this reduces to a simple three-site model~\cite{Dominguez.2016, Liu.2022, Tsintzis.2022}.
For simplicity, we treat the ABS as one pair of semiconducting states tunnel-coupled to the superconductor in the atomic limit~\cite{Bauer.2007}, simplifying the general expressions derived in Ref.~\cite{Liu.2022} (see Supplementary Information for details).
Andreev reflection at the semiconductor-superconductor interface hybridizes the two electronic states with even charge occupation, $\ket{0}$ and $\ket{2}$, with hybridization rate $\Gamma$.
The ground state of the ABS is a spin singlet of the form $\ket{\mathrm{S}} = u\ket{0}-v\ket{2}$, where $u,v>0$ are the normalized superposition coefficients determined by $\Gamma$ and $\mu$, the chemical potential of the electronic level before hybridization. 
Positive $\mu$ results in $u>v$ and negative $\mu$ leads to $u<v$~\cite{Bauer.2007}. 
The excited states of the ABS form a doublet $\DU,\DD$ where $\uparrow/\downarrow$ indicates, in the absence of spin-orbit coupling, the spin state of the single electron occupying the ABS (see the Supplementary Information for general spin-orbit-coupled scenarios).

Under zero external magnetic field, the doublet states are degenerate and the energy difference between $\ket{\mathrm{S}}$ and $\ket{\mathrm{D}}$ is $\EABS$, which reaches a minimum around $\mu=0$ (Fig.~\ref{fig: single_ABS}c)~\cite{Bauer.2007}.
An excitation from the ground state of the ABS to an excited state is said to be a Bogoliubov quasiparticle, having an electron-like part $u$ and a hole-like part $v$ in superposition.
The effective charge of the ABS is defined as the net charge character of this excitation, $-e(u^2-v^2)$, where $e>0$ is the elementary charge~\cite{Danon.2020,Menard.2020}.
This quantity ranges from $-e$ (electron-like) to $+e$ (hole-like).

We consider both CAR and ECT as coherent second-order processes that involve the virtual occupation of an ABS doublet as the intermediate state. 
ECT can take place through two paths. 
The first, marked in grey in Fig.~\ref{fig: single_ABS}a, involves the occupation of the ABS by adding an electron from one lead with a hopping amplitude proportional to $u$, followed by emptying of the ABS via ejection of the electron to the other lead, with an amplitude also proportional to $u$.  
The second, marked in green in Fig.~\ref{fig: single_ABS}a, occurs in the opposite order: an ABS is excited to $\ket{\mathrm{D}}$ by accepting a hole from one lead, with an amplitude proportional to $v$, and then relaxes to $\ket{\mathrm{S}}$ by ejecting a hole to the other lead, also with an amplitude proportional to $v$. 
As presented in Ref.~\cite{Liu.2022} and briefly here in Supplementary Information, these two paths interfere destructively due to fermion exchange statistics and the ECT-induced current, $\Iect$, is: 
\begin{equation}
I_\mathrm{ECT} = I_0 \abs{\frac{u^2-v^2}{\EABS/\Gamma}}^2
\end{equation}
where $I_0$ is a proportionality constant given by $I_0 = \frac{e}{\hbar} \cdot \frac{t^2_\mathrm{L} t^2_\mathrm{R} }{\Gamma^2 \gamma_\mathrm{DL}}$ and depends on the coupling between the QDs and the ABS ($t_\mathrm{L}$ and $t_\mathrm{R}$) as well as the lifetime of QDs due to coupling to the leads ($\gamma_\mathrm{DL}$) in the limit of electron temperature and tunnel couplings much smaller than bias voltage. 
Strikingly, the destructive interference results in a suppression of $\Iect$ near $\mu=0$ where $u^2=v^2=\frac{1}{{2}}$(Fig.~\ref{fig: single_ABS}e).

The process of CAR, depicted in Fig.~\ref{fig: single_ABS}b, can take place via two paths as well. 
In the first path (marked in green), an electron from the left lead populates the ABS with an amplitude proportional to $u$, followed by emptying of the ABS via accepting an electron from the right lead, with an amplitude proportional to $v$. 
In the second path, the roles of the left and right QDs are reversed. 
The two paths interfere constructively, yielding
\begin{equation}
    I_\mathrm{CAR} = I_0 \abs{\frac{2uv}{\EABS/\Gamma}}^2
\end{equation}
where $\Icar$ is the CAR-induced current, shown in Fig.~\ref{fig: single_ABS}e. 
The term $uv$ is significant only when $\abs{\mu}$ is small, leading to the peak in $\Icar$ around $\mu=0$  (Fig.~\ref{fig: single_ABS}e). 
This is also where ECT is diminished, allowing CAR to dominate over ECT.
Far away from ABS charge neutrality, ECT decays slower than CAR and becomes the dominant coupling mechanism, as it does not require electron-hole conversion to take place.
The distinct dependencies of CAR/ECT on $\mu$ thus enable us to tune the relative strengths between them via electrostatic gating.

To study our model experimentally, we focus on the range of $\VPG$ values between 0 and \SI{70}{mV} where a single ABS dominates the subgap spectrum (Fig.~\ref{fig: single_ABS}d). 
The ABS reaches a minimum around $\VPG = \SI{35}{mV}$ and merges with the superconducting gap below $\VPG = \SI{10}{mV}$ and above $\VPG = \SI{60}{mV}$. 
Fig.~\ref{fig: single_ABS}f shows $\Icar$ and $\Iect$ measured in the same $\VPG$ range with higher resolution in $\VPG$ than 
Fig.~\ref{fig: Correlation}e. As predicted, $\Icar$ features a narrow peak centered around the ABS energy minimum. 
$\Iect$ is non-zero in a wider range of $\VPG$ values and, as predicted, shows a dip when the ABS energy is minimal. 
We interpret this suppression as resulting from the destructive interference of the two ECT paths. 
We emphasize that this quantum mechanical interference is distinct from the cancellation between electron and hole currents as observed in three-terminal spectroscopy of hybrid nanowires~\cite{Menard.2020,Poschl.2022}. 
Note that, contrary to our theoretical model, $\Iect$ is not fully suppressed when $v > u$. 
This could be due to other ABSs at higher $\VPG$ that contribute to $\Iect$ or higher $\VPG$ increasing tunneling rates via gate cross-coupling. 
Similar observations of the $\VPG$ dependence reported here are reproduced in two more devices (Fig.~\ref{fig: devB} and Fig.~\ref{fig: devC}).


\begin{figure}
    \centering
    \includegraphics[width = 1\textwidth]{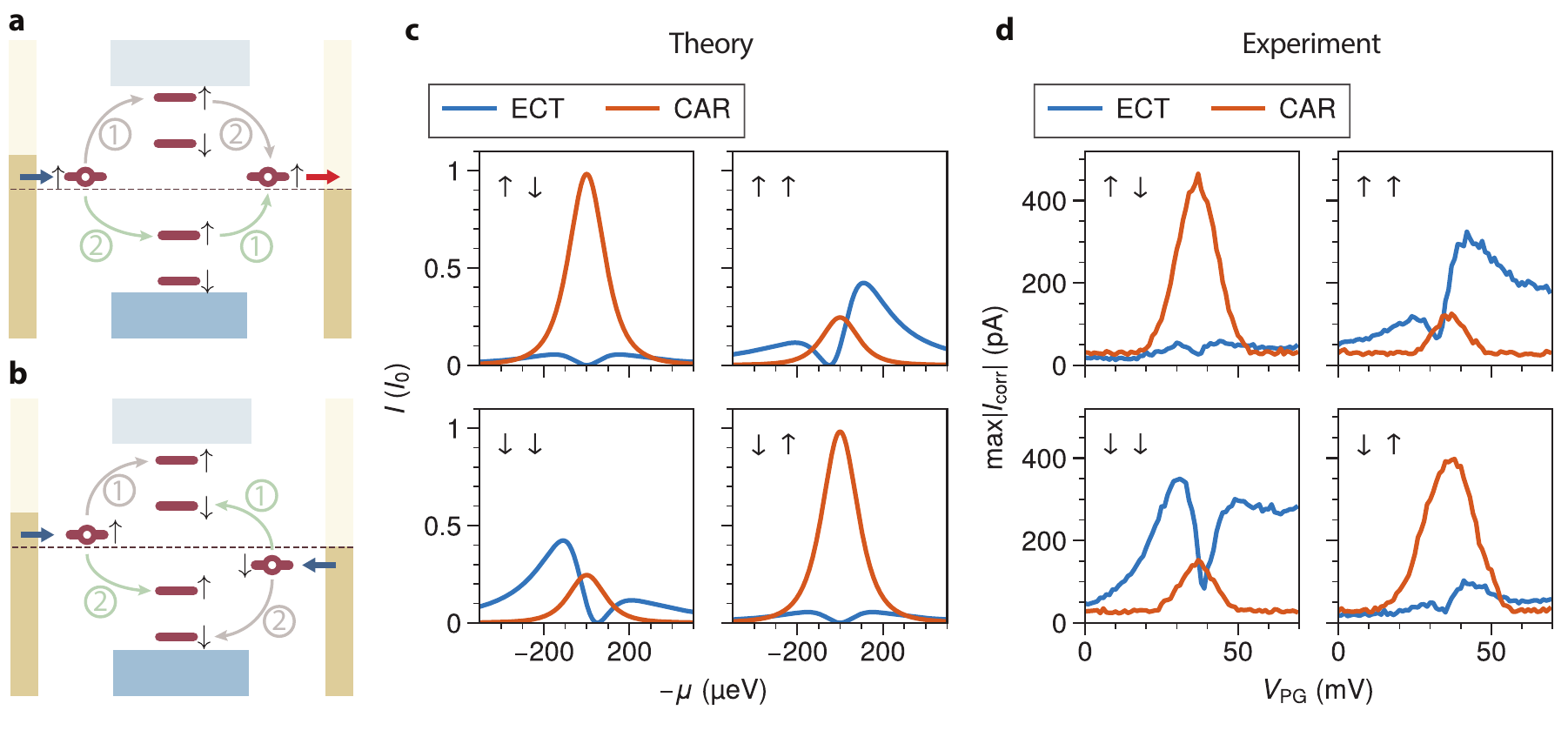}
    \caption{\textbf{CAR and ECT mediated by spin-polarized ABS. a.} ECT process mediated by a spin-polarized ABS between QDs in the $\uparrow \uparrow$ spin configuration. 
    \textbf{b.} CAR process mediated by spin-polarized ABS between QDs in the $\uparrow \downarrow$ spin configuration. 
    \textbf{c.} Calculation of the transmission probability of ECT and CAR via an atomic-limit ABS as a function $\mu$ at the four possible spin configurations of the QDs. Spin-orbit coupling is included in the calculation as a small spin-flipping factor ($\sigma = 0.2$) to allow for opposite-spin ECT and same-spin CAR (see Supplementary Information for model details). Other model parameters are $\Gamma = \SI{160}{\micro eV}$ and $E_\mathrm{Z} = \SI{100}{\micro eV}$ 
    \textbf{d.} A high-resolution CAR and ECT amplitudes while tuning $\VPG$ with $\vec{B} = \SI{80}{mT}$ (applied along the nanowire direction) at the four possible spin configurations of the QDs.}
    \label{fig: CAR_in_field}
\end{figure}


Application of a Zeeman field lifts the Kramers' degeneracy of the ABS and the QDs. 
The spin splitting of the QDs makes their charge transitions spin-polarized: the addition energy from $N$ to $N+1$ electrons becomes lower (spin-down, $\downarrow$), and that from $N+1$ to $N+2$ becomes higher (spin-up, $\uparrow$)~\cite{Hanson.2007}. 
We thus control the spins of the electrons participating in CAR and ECT by selecting the corresponding charge transitions~\cite{Wang.2022}. 
The odd states of the ABS split in energy, leading to two possible excitations from the ground state $\ket{\mathrm{S}}$: either to $\DD$ with an energy $E_\downarrow = \EABS - E_\mathrm{Z}/2$, or to $\DU$ with an energy $E_\uparrow = \EABS + E_\mathrm{Z}/2$, where $E_\mathrm{Z}$ is the Zeeman splitting of the ABS~\cite{Lee.2014}. 

Fig.~\ref{fig: CAR_in_field}a shows schematically the process of ECT in the presence of a Zeeman field when both QDs are tuned to the $\uparrow$ transition. 
Again, this process can take place via two paths. 
In the first path (marked in grey), an $\uparrow$ electron from one lead populates the  $\DU$ state of the ABS. 
Then the ABS is emptied by emitting an $\uparrow$ electron to the other lead through the QD. 
In the second process (marked in green), a hole from one lead hops into the ABS, exciting it into the $\DD$ state. 
The ABS then relaxes by emitting a hole to the other lead.  
The energies of the intermediate states in the two paths, $\DU$ and $\DD$, are split and the interference pattern is thus modified. 
The ECT-induced current is now of the form:
\begin{equation}
    \Iect^{\uparrow \uparrow} \propto \abs{  \frac{u^2}{E_\uparrow} - \frac{v^2}{E_\downarrow}  }^2
    \label{eq: IectB}
\end{equation}
Since $E_\downarrow<E_\uparrow$, ECT is stronger when the ABS is hole-like (large $v$) as seen in the $\uparrow\uparrow$ panel of Fig.~\ref{fig: CAR_in_field}c. 
Analogously, the ECT is higher when the ABS is electron-like ($u > v$) and both QDs are tuned to the $\downarrow$ transition.

CAR-induced currents are also modified by the Zeeman splitting of the ABS doublet state. 
CAR takes place in two paths involving both levels (shown schematically in Fig.~\ref{fig: CAR_in_field}b). 
In one path (marked in green), the ABS occupies the $\DD$ state by receiving a $\downarrow$ electron from one lead and is emptied by receiving an $\uparrow$ electron from the other lead. 
In the second path (marked in grey), the order is reversed and the ABS passes through the $\DU$ state. 
The probability for the CAR process is now:
\begin{equation}
    \Icar^{\uparrow \downarrow} \propto \abs{  \frac{uv}{E_\downarrow} + \frac{uv}{E_\uparrow}  }^2
    \label{eq: IcarB}
\end{equation}
This probability peaks at the ABS energy minimum, as seen in the relevant panel of Fig.~\ref{fig: CAR_in_field}c. Note that the expected CAR peak remains symmetric in $\mu$, in contrast to ECT.
Fig.~\ref{fig: CAR_in_field}d shows the measured $\Icar$ and $\Iect$ under the application of $\abs{\vec{B}} = \SI{80}{mT}$ along the nanowire direction, sufficient to fully spin-polarize the QDs ($E_\mathrm{Zeeman}^\mathrm{QD} \approx \SI{200}{\micro eV}$) and split the energy of the ABS ($\EZ \approx \SI{100}{\micro eV}$, see Fig.~\ref{ED_ABS_B}). 
Spin-orbit coupling in the nanowire allows for spin flipping processes --- equal-spin CAR and opposite-spin ECT --- to take place~\cite{Wang.2022}, allowing us to measure ECT and CAR in all possible spin configurations. 
$\Icar$ is symmetric around the ABS energy minimum and is generally larger for opposite-spin than equal-spin configurations. 
$\Iect$ in the $\uparrow \uparrow$ spin configuration is large when the ABS is hole-like ($v > u$) and is suppressed when it is electron-like (large $u$). 
The destructive interference dip is shifted from the ABS minimum towards lower $\VPG$. 
The opposite trend is observed in the $\downarrow \downarrow$ spin configuration: $\Iect$ is slightly larger when the ABS is electron-like, and the interference dip is shifted towards higher values of $\VPG$. 
$\Iect$ in the opposite-spins configuration is nearly symmetric around the ABS minimum and is generally suppressed with respect to $\Iect$ in the equal-spin configuration. 
Thus, all of the qualitative predictions of the model~\cite{Liu.2022} are verified in the measurements.


\begin{figure}[h]
    \centering
    \includegraphics[width = 1\textwidth]{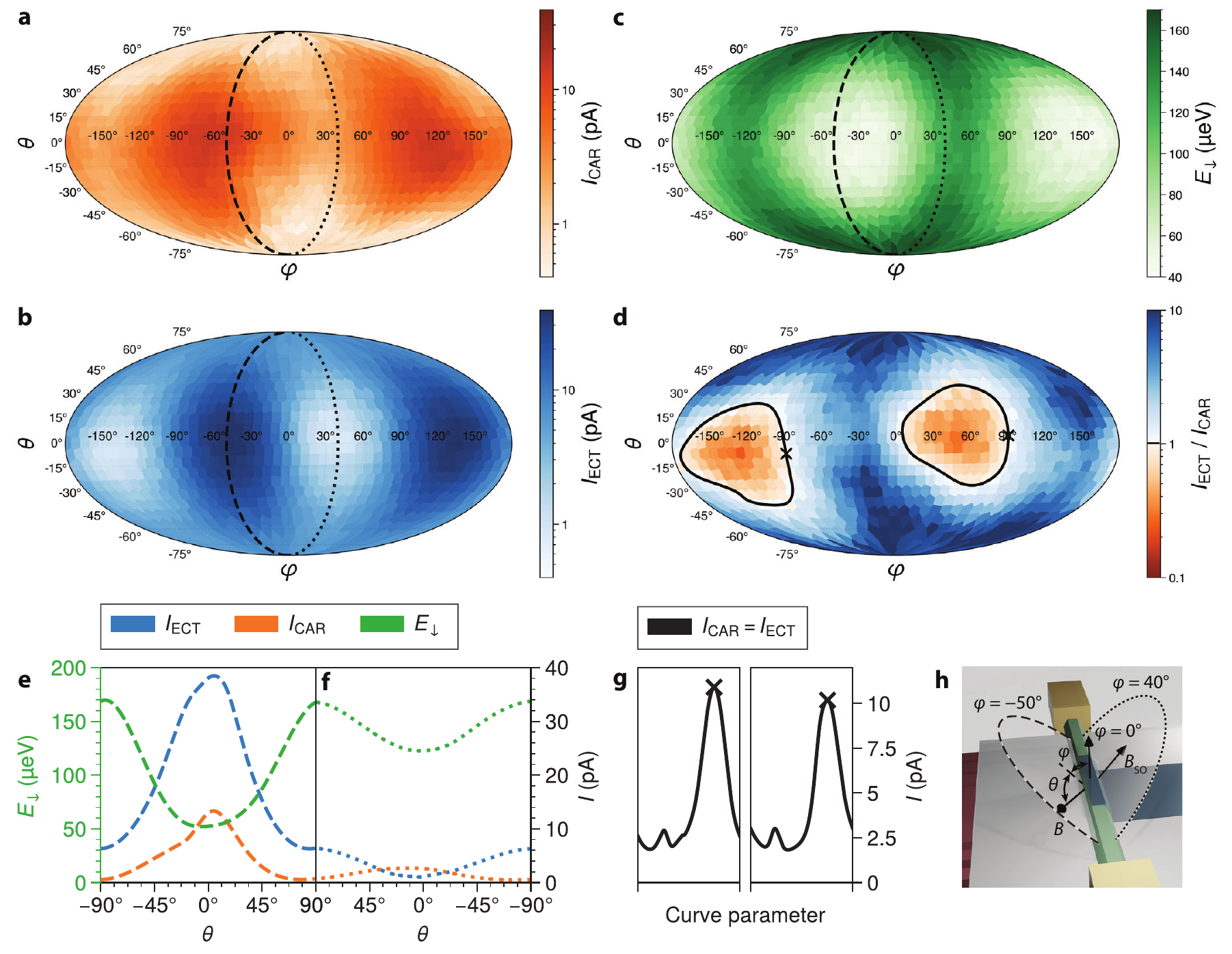}
    \caption{\textbf{Tuning CAR and ECT with magnetic field orientation.} \textbf{a-d.} Spherical plots: the center of every colored tile corresponds to a specific magnetic field orientation. Each panel is taken at fixed $\VPG$ and $\abs{\vec{B}}=\SI{80}{mT}$. $\VPG=\SI{475}{mV}$ in panel~c, while $\VPG=\SI{480}{mV}$ in a, b and d. The QD spin configuration is $\downarrow\uparrow$ for all panels. See Fig.~\ref{fig: globes-supp} for data corresponding to other spin configurations. 
    \textbf{a.} CAR-induced current as a function of magnetic field direction, extracted with the same method detailed in Fig.~\ref{fig: CAR-ECT} and used in the rest of the paper. 
    \textbf{b.} ECT-induced current as a function of magnetic field orientation. 
    \textbf{c.} Energy of the lowest-energy ABS extracted from local tunneling spectroscopy as a function of magnetic field orientation. 
    \textbf{d.} Ratio of the ECT and CAR currents from panels~b and a. Continuous lines highlight the locus of points where $\Icar = \Iect$; among them, the points with maximum current are marked with crosses. 
    \textbf{e.} Interpolation of data shown in panels a--c along the $\varphi = -50^\circ$ meridian. 
    \textbf{f.} Interpolation of data shown in panels a--c along the $\varphi = +40^\circ$ meridian. 
    \textbf{g.} $\Icar$ along the $\Icar = \Iect$ curves shown in panel~d. Negative-$\varphi$ points are parameterized and plotted on the left, positive-$\varphi$ points on the right. 
    \textbf{h.} Schematic defining $\theta$ and $\varphi$: $\theta = \pm 90^\circ$ is the direction parallel to the nanowire. $\theta = \varphi = 0^\circ$ is the direction perpendicular to the substrate.}
    \label{fig: globes}
\end{figure}


So far, we have discussed the dependence of CAR and ECT magnitudes as a function of the ABS charge at zero and finite Zeeman field. 
In the following, we report on the dependence of CAR and ECT on the direction of the applied magnetic field $\vec{B}$, at fixed $\VPG$. 
We measure a second device, B, with a longer superconducting segment ($\approx\SI{350}{nm}$, much larger than the superconducting coherence length in the Al film) and no Pt layer on top of the Al.
The schematic in Fig.~\ref{fig: globes}h indicates the angles $\theta$ and $\varphi$ defining the field direction of $\vec{B}$. 
The QDs are set to the $\downarrow\uparrow$ spin configuration and $\VPG$ is selected such that ECT is stronger than CAR when the field is parallel to the nanowire. 
In Fig.~\ref{fig: globes}, panels a and b show $\Icar$ and $\Iect$ when the angle of $\vec{B}$ is varied over a sphere.
Panel~c shows the energy of the lowest ABS at a similar $\VPG$ (see Methods and Fig.~\ref{fig: ED_gapsize_spec} for analysis details). 
All three quantities are anisotropic and CAR and ECT amplitudes are overall negatively correlated to $E_\downarrow$ across the plotted globes, as expected for virtual tunneling processes.
Below, we examine the rotational dependence of CAR and ECT along two exemplary meridians of the globe (dashed and dotted lines in panels a to c) in order to separate anisotropy due to ABS energy from that caused by spin.

As discussed above, CAR and ECT amplitudes are inversely proportional to the ABS energy. 
This effect is highly visible in Fig.~\ref{fig: globes}e, where we plot $E_\downarrow$, $\Icar$ and $\Iect$ along the meridian with $\varphi = -50^\circ$ (dashed line in Fig.~\ref{fig: globes}a-c). 
Here, $E_\downarrow$ is significantly modulated between $\sim \SI{170}{\micro eV}$ and $\sim \SI{50}{\micro eV}$ and, accordingly, both $\Icar$ and $\Iect$ increase drastically at the energy minimum.
In contrast, very different pattern can be obtained when we rotate the magnetic field along the meridian of $\varphi = 40^\circ$ (dotted line in Fig.~\ref{fig: globes}a-c). 
Fig.~\ref{fig: globes}f shows that, along this meridian, $E_\downarrow$ changes by a small amount. 
As before, $\Icar$ is enhanced where $E_\downarrow$ is minimal.
However, $\Iect$ varies in the opposite way and becomes completely suppressed around $\theta=0$ (perpendicular to the nanowire axis).  
This suppression is generic across various $\VPG$ values and therefore not explained by either the energy or the charge of the ABS.
We attribute the reduction of opposite-spin ECT along this specific direction to spin blockade~\cite{Wang.2022}.
When the QDs select opposite spins, spin precession due to spin-orbit coupling enables the presence of some $\Iect$~\cite{Liu.2022}. 
However, if the applied $\vec{B}$ is parallel to the effective spin-orbit field $\BSO$, no spin precession occurs and therefore ECT is suppressed between QDs with opposite spins~\cite{Hofmann.2017, Wang.2018}. 
The observation of this type of spin blockade reveals the orientation of the spin-orbit field. 
Compared to prior works measuring the spin-orbit field direction in hybrid nanowires via superconducting gap size anisotropy, the method presented here using spin conservation to detect $\BSO$ direction is less prone to other effects such as orbital depairing and $g$-factor anisotropy (Fig.~\ref{fig: ABS-globes-supp}).

With these two effects in mind, we summarize the angle dependence of CAR and ECT over the entire sphere as follows.
First, there exists one special $\vec{B}$ direction along which equal-spin CAR and opposite-spin ECT are strongly suppressed (see Fig.~\ref{fig: globes-supp} for other spin combinations).
We interpret this as a spin-blockade effect and its direction as that of the spin-orbit field.
Away from this blockaded direction, multiple factors compete to influence the amplitudes of CAR and ECT, such as the angle between $\vec{B}$ and $\BSO$ and the energy of the mediating ABS.

This combination of anisotropic ABS energy and spin-orbit coupling makes the $\vec{B}$ direction dependence of CAR and ECT very rich, enabling further tuning of their relative amplitudes.
Fig.~\ref{fig: globes}d shows the ratio between $\Iect$ and $\Icar$ as a function of $\vec{B}$ orientation. 
Here, due to the aforementioned influence of the ABS charge, $\Iect$ is larger than $\Icar$ on most of the sphere. 
However, since ECT is suppressed along a specific direction, the ratio between $\Iect$ and $\Icar$ can be inverted. 
Such tunability allows for $\Icar = \Iect$, the sweet spot essential for the realization of Poor Man's Majoranas in a minimal Kitaev chain~\cite{Leijnse.2012, Dvir.2022}. 
Fig.~\ref{fig: globes}d shows with continuous lines the locus of points where $\frac{\Iect}{\Icar}=1$ and Fig.~\ref{fig: globes}f reports the corresponding current values, highlighting with crosses the points where $\Icar (=\Iect)$ is maximal. 
It is therefore evident that the $\vec{B}$ dependence of CAR and ECT not only enables the tuning to the $\Icar=\Iect$ sweet spot, but also allows optimization of their strengths.

\vskip\baselineskip

In summary, we have measured ECT- and CAR-induced currents mediated by ABSs formed in a proximitized InSb nanowire. 
We show that the amplitudes of both processes depend on the charge of the ABSs, and are thus highly tunable via electrostatic gating. Particularly, we show that ECT is significantly suppressed when the ABS is charge-neutral due to destructive interference originating from fermionic exchange statistics. 
Furthermore, we examine how the interference pattern and the balance between ECT and CAR is shifted when the applied magnetic field spin-polarizes the QDs and splits the energy of the ABS. 
Finally, we measure how the magnetic field orientation modifies both the energy of the ABS and the effect of spin-orbit coupling, adding another independent knob to tune CAR and ECT. 
These results demonstrate deterministic control of the relative amplitudes of CAR and ECT, forming the foundation of realizing an artificial Kitaev chain~\cite{Dvir.2022}. 


\newpage

\section{Data Availability and Code Availability}

Raw data presented in this work, the data processing/plotting code, and the code used for the theory calculations are available at \url{https://doi.org/10.5281/zenodo.7395016}.

\section{Acknowledgments}

This work has been supported by the Dutch Organization for Scientific Research (NWO) and Microsoft Corporation Station Q.
We thank Gijs de Lange for helpful discussions.

\section{Author contributions}
GW, GPM, NvL, AB, FZ and DvD fabricated the devices. GW, TD, SLDtH, AB and DX performed the electrical measurements. TD and GW designed the experiment. AB, GW, and TD analyzed the data. AB, TD, and LPK prepared the manuscript with input from all authors. TD and LPK supervised the project. CXL developed the theoretical model with input from MW. SG, GB and EPAMB performed InSb nanowire growth.

\bibliography{bibliography}

\pagebreak

\section{Supplementary Information}

\setcounter{equation}{0}

\subsection{Theoretical model}
In this section, we show analytically how to obtain the CAR/ECT couplings/currents based on our theoretical model. The calculation is based on Ref.~\cite{Liu.2022}, but specialized to the atomic limit.

The model Hamiltonian for the dot-hybrid-dot system is
\begin{equation}
    H=H_{\text{dot}} + H_{\text{hybrid}} + H_{\text{tunnel}}
\end{equation}
We first introduce the Hamiltonian for two quantum dots, $H_{\text{dot}}$:
\begin{equation}
    H_{\text{dot}} = \varepsilon_L d\dg_{L\eta} d_{L\eta} + 
    \varepsilon_R d\dg_{R\sigma} d_{R\sigma}
\end{equation}
Here, $\varepsilon_{L/R}$ are the dot energies relative to the Fermi energy, and $d_{L\eta}, d_{R\sigma}$ denote the spin-polarized dot levels in the presence of a large magnetic field. 
Note that no summation is taken over the spin indices $\eta$ and $\sigma$.

Next, $H_{\text{hybrid}}$ describes the hybrid segment with two Andreev bound states. 
In general, we can write down its Hamiltonian in diagonalized form using Bogoliubov quasiparticle operators $\gamma_+$ and $\gamma_-$:
\begin{equation}
    H_{\text{hybrid}} = E_+ \gamma\dg_+ \gamma_+ + E_- \gamma\dg_- \gamma_-
\end{equation}
More concretely, we consider a superconducting-atomic-limit model~\cite{Bauer.2007} under weak spin-orbit coupling compared to Zeeman field.
This means we use the pseudospin labels $+,-$ instead of $\uparrow,\downarrow$ but treat the spin-splitting between the two as approximately $\EZ$.
We also set charging energy of the ABS to zero, as it is strongly screened by the grounded Al film in the experiment.
Using electron annihilation operators $a_+,a_-$, the atomic-limit Hamiltonian is 
\begin{equation}
    H_\mathrm{hybrid} \approx \mu \left(a_+^\dagger a_+ + a_-^\dagger a_- \right) + \frac{\EZ}{2} \left(a_+^\dagger a_+ - a_-^\dagger a_- \right) + \Gamma a_+^\dagger a_-^\dagger + \hc
\end{equation}
This model can be solved exactly. The ABS energies are
\begin{align}
    E_{\pm} \approx \sqrt{\mu^2 + \Gamma^2} \pm \frac{\EZ}{2},
\end{align}
where $\mu$ is the chemical potential in the hybrid segment which is controlled by the plunger gate voltage $\VPG$ in the experimental device, $\Gamma$ is the superconducting coupling strength and $\EZ$ is the Zeeman spin splitting.
The corresponding wavefunctions of the ABSs are
\begin{align}
    \gamma_+ = u a_+ + va\dg_-, \quad \gamma_- = v a_+ - u a_-,
    \label{eq:ABS_wavefunction}
\end{align}
where $u^2 = 1 - v^2 = \frac12 \left(1 + \frac{\mu}{\sqrt{\mu^2 + \Gamma^2}}\right)$ are the BCS coherence factors characterizing the electron and hole components of the ABSs.

Finally, $H_{\text{tunnel}}$ is the tunnel Hamiltonian between the dots and the hybrid segment:
\begin{align}
    H_{\text{tunnel}} = \sum_{a=L,R} &t_a \left[ \left( \cos\theta_a a\dg_+ - \sin \theta_a a\dg_{-} \right) d_{a\uparrow} 
    + \left( \sin \theta_a a\dg_+ + \cos \theta_a a\dg_-  \right) d_{a\downarrow} \right] \nn
  &+ \hc
\end{align}
Here we only consider the scenario where the spin-orbit field ($\propto \sigma_y$) is perpendicular to the globally applied magnetic field ($\propto \sigma_z$).
In particular, $\theta_R = -\theta_L = k_{so}L/2$ describes the spin precession in the hybrid region due to spin-orbit interaction, where $k_{so}=m\alpha_R/\hbar^2$ is the spin-orbit wave-vector and $L$ is the length of the hybrid segment.
In the weak spin-orbit interaction regime ($k_{so}L/2 \ll 1$), the ABS of $\gamma_+$ is mainly spin-up, but it can also accommodates spin-down QD electrons with amplitude $\sim \sin(k_{so}L/2)$. Similar pictures hold for $\gamma_-$ as well.
Note that although we include both $\uparrow$ and $\downarrow$ in $H_{\text{tunnel}}$ for dots, in the calculation, we only include one spin species depending on the choice of $\eta$ and $\sigma$ in $H_{\text{dot}}$, as the experimental measurements are spin-polarized.

In the tunneling regime $t_{L/R} \ll \Gamma$, the effective Hamiltonian of the coupled quantum dots can be obtained using the perturbation theory as below
\begin{align}
    H_{\text{eff}} &= H_{\text{dot}} - H_{\text{tunnel}}\frac{1}{H_{\text{hybrid}}} H_{\text{tunnel}} + O(H^3_{\text{tunnel}}) \nn
    &\approx H_{\text{dot}} - \Gamma^{\text{ECT}}_{\eta \sigma} d\dg_{L\eta} d_{R\sigma}
    - \Gamma^{\text{CAR}}_{\eta \sigma} d_{L\eta} d_{R\sigma} + h.c.,
    \label{eq:perturbation_theory}
\end{align}
where $\Gamma^{\text{CAR}}_{\eta \sigma}$ and $\Gamma^{\text{ECT}}_{\eta \sigma}$ are the spin-selective CAR and ECT couplings between quantum dots.
Interestingly, as shown in Ref.~\cite{Liu.2022} and explained in the main text of this work, the strengths of the effective couplings can be extracted from the resonant current measured in a three-terminal setup, that is 
\begin{align}
    I^{\text{max}}_{\text{CAR/ECT}} \propto \abs{\Gamma^{\text{CAR/ECT}}_{\eta \sigma}}^2.
\end{align}
In the $\uparrow \uparrow$ channel, from Eq.~\eqref{eq:perturbation_theory}, the CAR coupling is 
\begin{align}
    \Gamma_{\uparrow \uparrow}^{\text{CAR}} /t_Lt_R &=
- \left( -\sin \theta_L \cos \theta_R \frac{uv}{E_+} -\sin \theta_R \cos \theta_L \frac{(-u)v}{E_-} \right) \nn
& + \left(  -\sin \theta_R \cos \theta_L \frac{uv}{E_+} -\sin \theta_L \cos \theta_R \frac{(-u)v}{E_-} \right) \nn
&= - \sin ( \theta_R - \theta_L ) \left( \frac{uv}{E_+} + \frac{uv}{E_-} \right) \nn
&= - \sin ( k_{so}L ) \left( \frac{uv}{E_+} + \frac{uv}{E_-} \right),
\end{align}
which gives 
\begin{align}
    I_{\uparrow \uparrow}^{\text{CAR}} = I_0 \cdot \sigma \cdot \abs{ \frac{uv}{E_+} + \frac{uv}{E_-}  }^2,
    \label{eq:I_CAR_uu}
\end{align}
where $\sigma = \sin^2 ( k_{so}L )$ is the flipping rate due to spin-orbit interaction, and $I_0 = e t^2_L t^2_R /(\hbar \Gamma^2 \gamma_{DL})$.
Similarly, 
\begin{align}
\Gamma_{\uparrow\uparrow}^{\text{ECT}}/t_Lt_R &= 
\left( \cos \theta_L \cos \theta_R \frac{u^2}{E_+} + \sin \theta_L \sin \theta_R \frac{(-u)^2}{E_-} \right) \nn
&- \left( \cos \theta_L \cos \theta_R \frac{v^2}{E_-} + \sin \theta_L \sin \theta_R \frac{v^2}{E_+}   \right)\nn
&= \cos (k_{so}L) \left(  \frac{u^2}{E_+} - \frac{v^2}{E_-} \right)
- \sin^2 (k_{so}L/2) \frac{2\EZ}{E_+ E_-} \nn
\end{align}
giving 
\begin{align}
    I_{\uparrow \uparrow}^{\text{ECT}} \approx I_0 \cdot (1-\sigma) \cdot \abs{  \frac{u^2}{E_+} - \frac{v^2}{E_-}  }^2.
    \label{eq:I_ECT_uu}
\end{align}
Here, the ECT current in Eq.~\eqref{eq:I_ECT_uu} shows a destructive interference between two virtual paths.
In the first path, an electron first hops from the right dot into the hybrid before it hops out to the left ($\propto u^2$). 
In the second path, an electron first escapes from the hybrid segment to the left dot, leaving behind a hole-like ABS excitation which is later annihilated when a second electron jumping in from the right dot ($\propto -v^2$).
The minus sign responsible for the destructive interference stems from fermionic statistics when switching the order of two hopping events.
In contrast, the CAR current in Eq.~\eqref{eq:I_CAR_uu} shows a constructive interference pattern.
Because the CAR process is proportional to $ uv$ instead of $u^2$ or $v^2$, an additional minus sign in the ABS wavefunctions (see Eq.~\eqref{eq:ABS_wavefunction}) cancels the minus sign from fermionic statistics, yielding a constructive interference between the two virtual paths.
A similar analysis and calculation gives the current in the $\uparrow \downarrow$ channel:
\begin{align}
    & I_{\uparrow \downarrow}^{\text{CAR}} = I_0 \cdot (1-\sigma) \cdot \abs{ \frac{uv}{E_+} + \frac{uv}{E_-}  }^2, \nn
    & I_{\uparrow \downarrow}^{\text{ECT}} = I_0 \cdot \sigma \cdot \abs{\frac{\mu}{E_+E_-}}^2.
\end{align}
The currents in the remaining channels are readily obtained using the following symmetry relation~\cite{Liu.2022}:
\begin{align}
    &I_{\downarrow \uparrow}^{\text{CAR}}(\EZ) = I_{\uparrow \downarrow}^{\text{CAR}}(\EZ), \nn
    &I_{\downarrow \uparrow}^{\text{ECT}}(\EZ) = I_{\uparrow \downarrow}^{\text{ECT}}(\EZ),\nn
    &I_{\downarrow \downarrow}^{\text{CAR}}(\EZ) = I_{\uparrow \uparrow}^{\text{CAR}}(\EZ), \nn
    &I_{\downarrow \downarrow}^{\text{ECT}}(\EZ) = I_{\uparrow \uparrow}^{\text{ECT}}(-\EZ).
\end{align}
In the absence of Zeeman field and when the dot occupancy is tuned at the transition of $N$ to $N+1$, currents from all the four spin channels are allowed, yielding the total current 
\begin{align}
    &I^{\text{CAR}} = \sum_{\eta, \sigma = \uparrow, \downarrow} I_{\eta \sigma}^{\text{CAR}}(\EZ=0) = 2 \cdot I_0 \cdot \abs{\frac{2uv}{E}}^2, \nn
    &I^{\text{ECT}} = \sum_{\eta, \sigma = \uparrow, \downarrow} I_{\eta \sigma}^{\text{ECT}}(\EZ=0) = 2 \cdot I_0 \cdot \abs{\frac{u^2 - v^2}{E}}^2,
\end{align}
where $E = \sqrt{\mu^2 + \Gamma^2}$.

\pagebreak

\subsection{Methods}

\subsubsection{Device fabrication}
Fig.~\ref{fig: Correlation}c shows a device schematic and the electrical circuit used to measure it. Scanning electron microscope images of reported devices are shown in Fig.~\ref{fig: SEMs}. 
For device A, InSb nanowires were deposited on pre-fabricated metallic gates, separated from the nanowire by a $\SI{20}{nm}$ layer of HfO$_2$ dielectric. 
Using the shadow lithography technique~\cite{Heedt:2021_NC, Borsoi:2021_AFM}, an $\SI{8}{nm}$ layer of Al was deposited on top of the middle segment of the nanowire, followed by a $\SI{2}{\angstrom}$ layer of Pt and an additional $\SI{20}{nm}$ of AlO$_x$ capping layer. 
Normal Cr/Au contacts were later fabricated using standard e-beam lithography. Full details of the fabrication can be found in the Supplementary Information of Ref.~\cite{Mazur.2022}. 
For devices B and C, no additional Pt layer was used. 
For device C only, a double dielectric layer was used: $\SI{10}{nm}$ of Al$_2$O$_3$ followed by $\SI{10}{nm}$ of HfO$_2$.

\subsubsection{Data processing}
Transport was measured by applying DC voltage biases on the left and the right leads ($\VL,\VR$) and measuring the resulting DC currents on both sides ($\IL, \IR$). 
Local ($\GLL = dI_\mathrm{L}/dV_\mathrm{L}$, $\GRR = dI_\mathrm{R}/dV_\mathrm{R}$) and nonlocal ($\GRL  = dI_\mathrm{R}/dV_\mathrm{L}, \GLR  = dI_\mathrm{L}/dV_\mathrm{R}$) conductances were obtained as numerical derivatives of the DC currents after applying a Savitzky-Golay filter, unless otherwise specified. 
The $E_\downarrow$ energy of Fig.~\ref{fig: globes}c and $E_\textrm{gap}$ values of Fig.~\ref{fig: ABS-globes-supp} are extracted from $\IR(\VR)$ tunnel spectroscopy measurements by detecting where $\abs{\IR}$ exceeds a 5\% threshold of its value far outside the superconducting gap (see data repository for details and Fig.~\ref{fig: ED_gapsize_spec} for comparison between conductance spectroscopy and $E_\downarrow$ thus extracted). 
All measurements were conducted in a dilution refrigerator with a measured electron temperature of $\sim \SI{50}{mK}$. 

\subsubsection{Extraction of CAR and ECT amplitudes}
ECT-induced currents are measured with fixed voltage biases such that $\VL \neq \VR$.
Due to energy conservation, ECT-induced currents arise when $\VLD$ and $\VRD$ fulfill the condition that the chemical potentials of both QDs are aligned and within the bias window (shown schematically in Fig.~\ref{fig: CAR-ECT}b). ECT-induced current is detected as correlated current flowing from one lead to the other.
Similarly, CAR-induced currents are measured with fixed voltage biases such that $\VL \neq -\VR$. CAR-induced current arise when the QD chemical potentials are equal in magnitude with an opposite sign with respect to the Fermi energy (shown schematically in Fig.~\ref{fig: CAR-ECT}a). CAR currents flow jointly from the leads to the superconductor or vice-versa.
For each CAR and ECT measurement, $\VLD$ and $\VRD$ are swept around a charge degeneracy point of each dot. We measure $\IL$ and $\IR$ and calculate the correlated current $\Icor \equiv \mathrm{sign} \left( \IL \IR \right) \sqrt{ \abs{ \IL \IR }}$.
In this manuscript, as in~\cite{Wang.2022}, the maximum of $\Icor$ is taken as a proxy of the CAR strength $\Icar\equiv\max(\Icor)$, and minus the minimum of $\Icor$ is taken as a proxy of the ECT strength $\Iect\equiv-\min(\Icor)$. Notice that in the absence of CAR or ECT signal, $\max(\Icor)$ and $-\min(\Icor)$ return the background noise extrema. Background noise level is $\sim \SI{30}{pA}$ for device A and $\sim \SI{1}{pA}$ for devices B and C. We note that, instead of taking the bare maximum and minimum of $\Icor$, averaging procedures can improve the signal-to-noise ratio~\cite{Wang.2022.2DEG}, although come at the price of having to set an arbitrary threshold.
Every CAR data point in Fig.~\ref{fig: Correlation}e,~\ref{fig: single_ABS}f,~\ref{fig: CAR_in_field}d is taken from a $\VLD$-$\VRD$ sweep with symmetric biases, while every ECT data point is taken from a subsequent $\VLD$-$\VRD$ sweep with anti-symmetric biases.
Every data point of Fig.~\ref{fig: globes}a-b comes from a single $\VLD$-$\VRD$ sweep with finite $\VL > 0$ while $\VR=0$; in this case, positive $\muL$ allows for CAR and negative $\muL$ allows for ECT.
For every CAR and ECT measurement, we make sure that the bias voltages $\VL$ and $\VR$ are smaller than the ABS energy.
$\Icar$, $\Iect$ and $E_\downarrow$ values along specific lines shown in Fig.~\ref{fig: globes}e-f are extracted from a spherical interpolation of Fig.~\ref{fig: globes}a-c data. The interpolation is performed using the \texttt{scipy} implementation of a smooth bivariate spline approximation in spherical coordinates.
Code generating all plots is available in the linked repository.

\pagebreak

\section{Extended Data}
\renewcommand\thefigure{ED\arabic{figure}}
\setcounter{figure}{0}
\begin{figure}[h]
    \centering
    \includegraphics[width=0.9\textwidth]{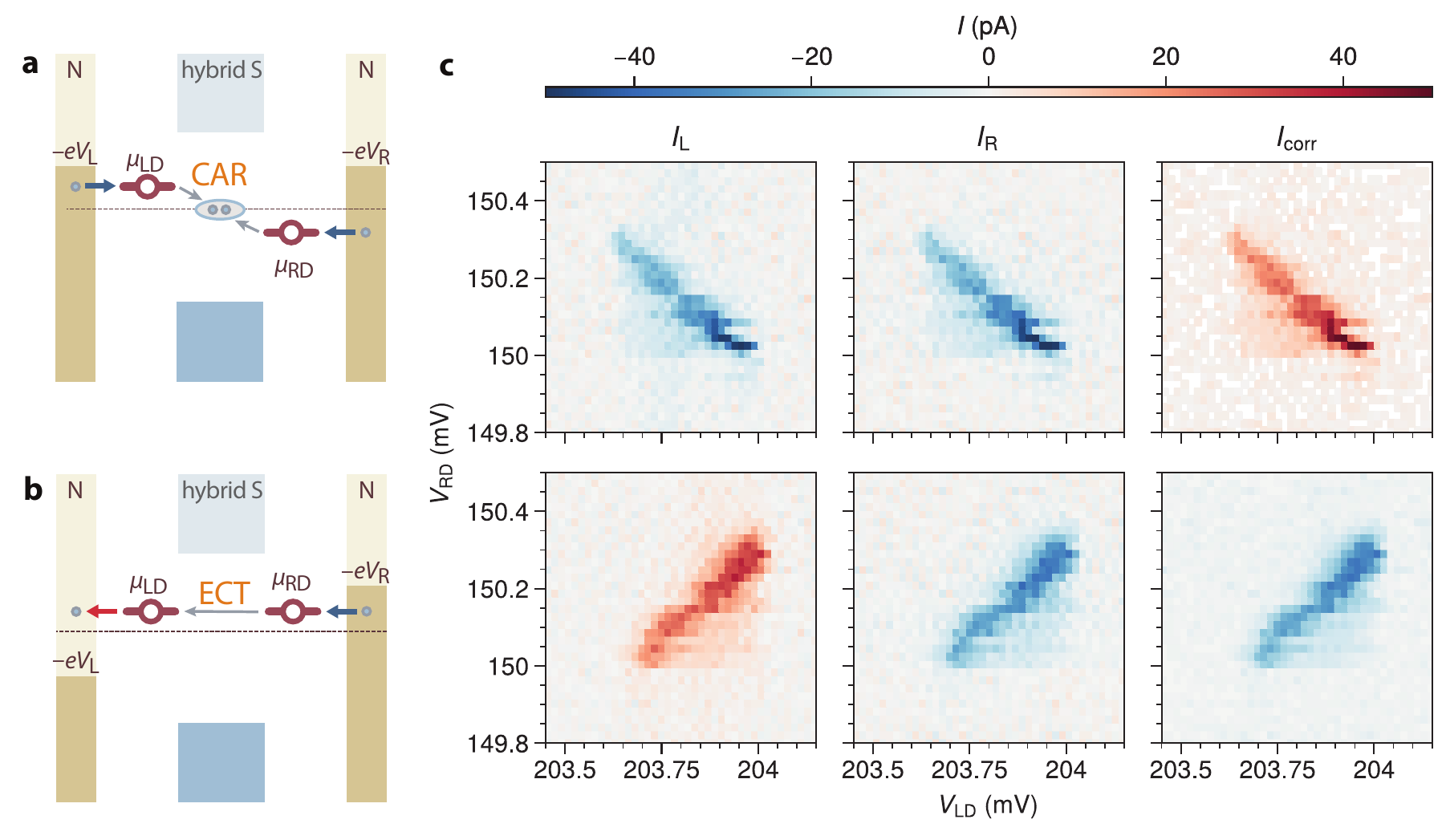}
    \caption{\textbf{Extraction of CAR and ECT amplitudes from measurements.}
    \textbf{a.} Schematic depiction of the CAR process: a Cooper pair splits into two electrons, one drained to the left lead via the left QD and one drained to the right lead via the right QD. 
    \textbf{b.} Schematic depiction of the ECT process: an electron goes from the left lead to the hybrid via the left QD and it is drained to the right lead via the right QD.  
    \textbf{c.} Measurement of the CAR-induced current (first row) and the ECT-induced current (second row), as in Ref.~\cite{Wang.2022}, around a charge degeneracy point. The right column shows the correlated current: $\Icor = \mathrm{sign} \left( \IL \IR \right) \sqrt{ \abs{ \IL \IR }}$. CAR-induced current is recognizable by positive $\Icor$ and by the anti-diagonal slope of the charge degeneracy point, due to $\muL = -\muR$. ECT-induced current is recognizable by negative $\Icor$ and by the diagonal slope of the charge degeneracy point, due to $\muL = \muR$.
    If the biases are symmetric ($\VL = \VR$), then current cannot flow from one lead to the other and therefore ECT-induced current is absent. If the biases are anti-symmetric ($\VL = -\VR$), then current cannot be drained by the superconducting ground from both normal leads and therefore CAR-induced current is absent. If the biases are neither symmetric nor anti-symmetric, then both CAR- and ECT-induced currents may be present in one sweep. Finally, CAR and ECT amplitudes are extracted as $\Icar\equiv\max(\Icor)$ and $\Iect\equiv-\min(\Icor)$.
    }
    \label{fig: CAR-ECT}
\end{figure}

\begin{figure}[htbp]
\centering
\includegraphics[width=0.9\textwidth]{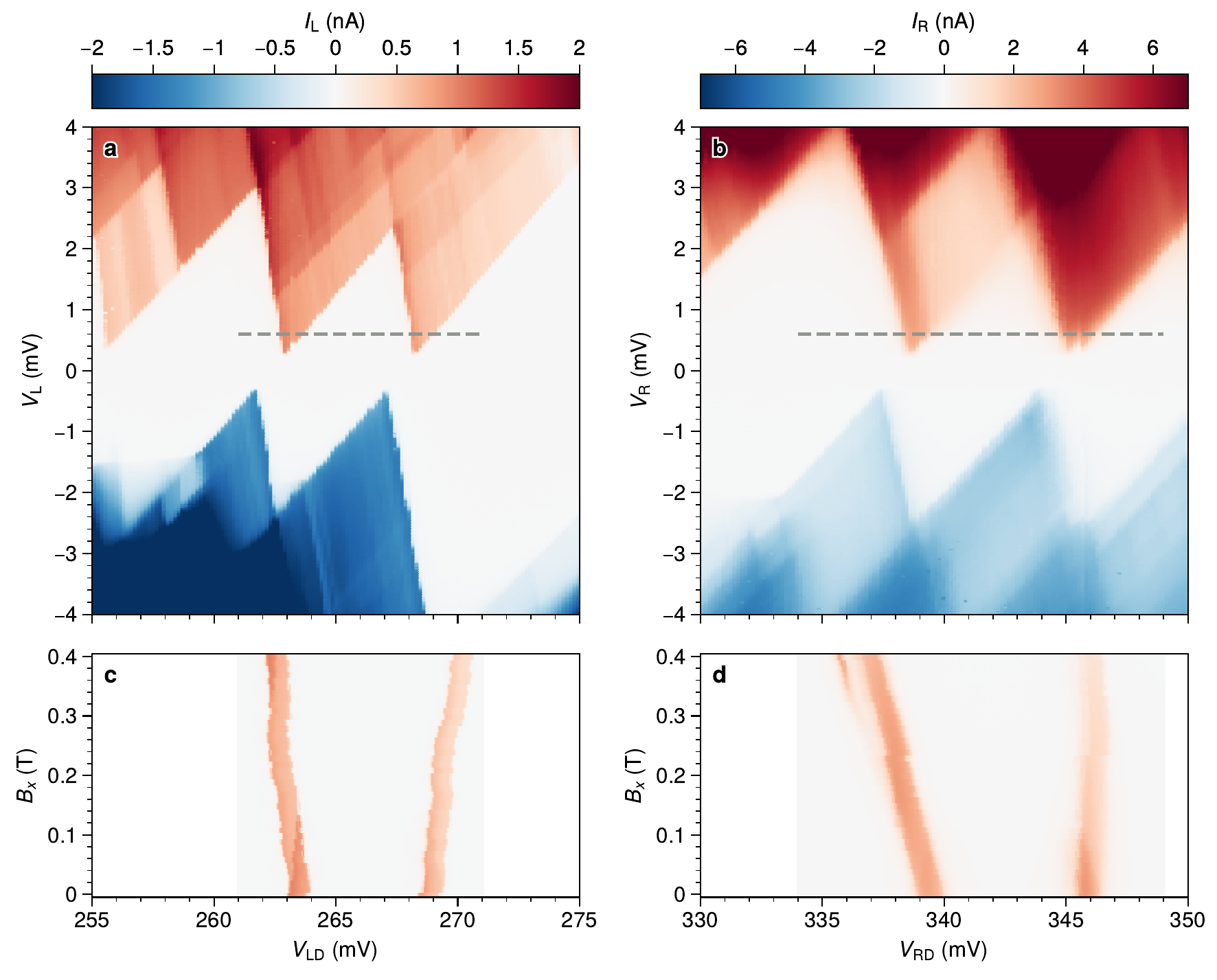}
\caption{\textbf{QD characterization in device~A}.
\textbf{a.} Coulomb blockade diamonds of the left QD, from which we estimate the charging energy to be $E_c = \SI{2.15}{meV}$ and lever arm $\alpha = 0.4$.
\textbf{b.} Coulomb blockade diamonds of the right QD. We estimate $E_c = \SI{2.3}{meV}$ and $\alpha = 0.35$. 
In both QDs, no sub-gap current is visible, indicating QDs are weakly coupled to S and retain their charge eigenstates.
Dashed lines highlight a $\SI{600}{\micro V}$ voltage bias set for panels c and d.
\textbf{c.} Current through the left QD at $\VL=\SI{600}{\micro V}$ measured against gate voltage and magnetic field along the nanowire, $B_x$.
Spin-degenerate orbitals Zeeman-split in opposite directions. We estimate a $g$-factor of $g=40$. 
\textbf{d.} Current through the right QD at $\VR=\SI{600}{\micro V}$. $g = 46$.
}\label{ED_dotchar}
\end{figure}

\begin{figure}[htbp]
\centering
\includegraphics[width=0.5\textwidth]{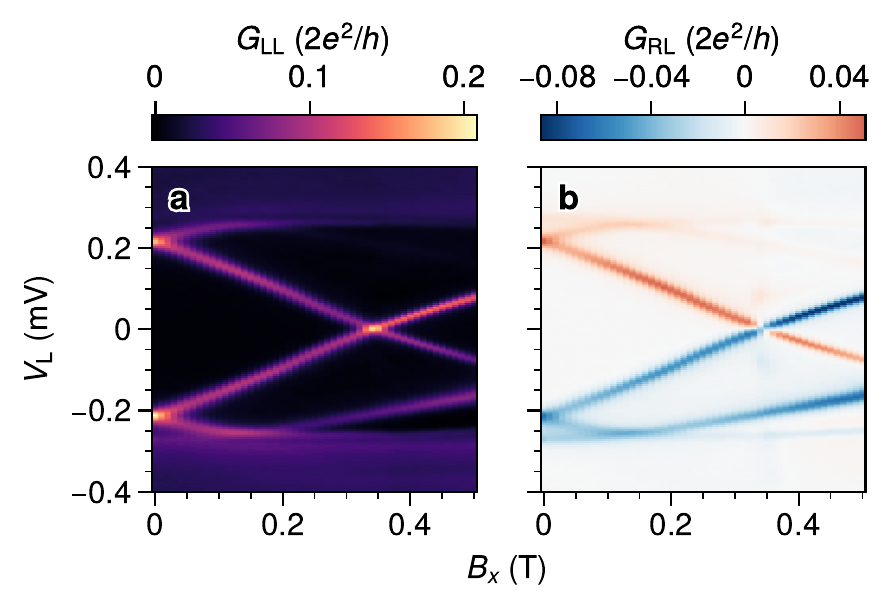}
\caption{\textbf{$\vec{B}$ dependence of the energy spectrum in the middle hybrid segment of Device~A revealing an ABS.}
\textbf{a.} Local conductance $G_\mathrm{LL} \equiv \mathrm{d}\IL / \mathrm{d} \VL$ measured with standard lock-in techniques.
The $g$-factor of the superconducting-semiconducting hybrid state is seen to be 21 from this plot, smaller than that in QDs.
\textbf{b.} Nonlocal conductance $G_\mathrm{RL}\equiv \mathrm{d}\IR / \mathrm{d} \VL$. The presence of nonlocal conductance corresponding to this state proves this is an extended ABS residing under the entire hybrid segment, tunnel-coupled to both sides.
We note that this is the same dataset presented in other manuscripts~\cite{Mazur.2022, Wang.2022}.
}\label{ED_ABS_B}
\end{figure}

\begin{figure}
    \centering
    \includegraphics[width = 1\textwidth]{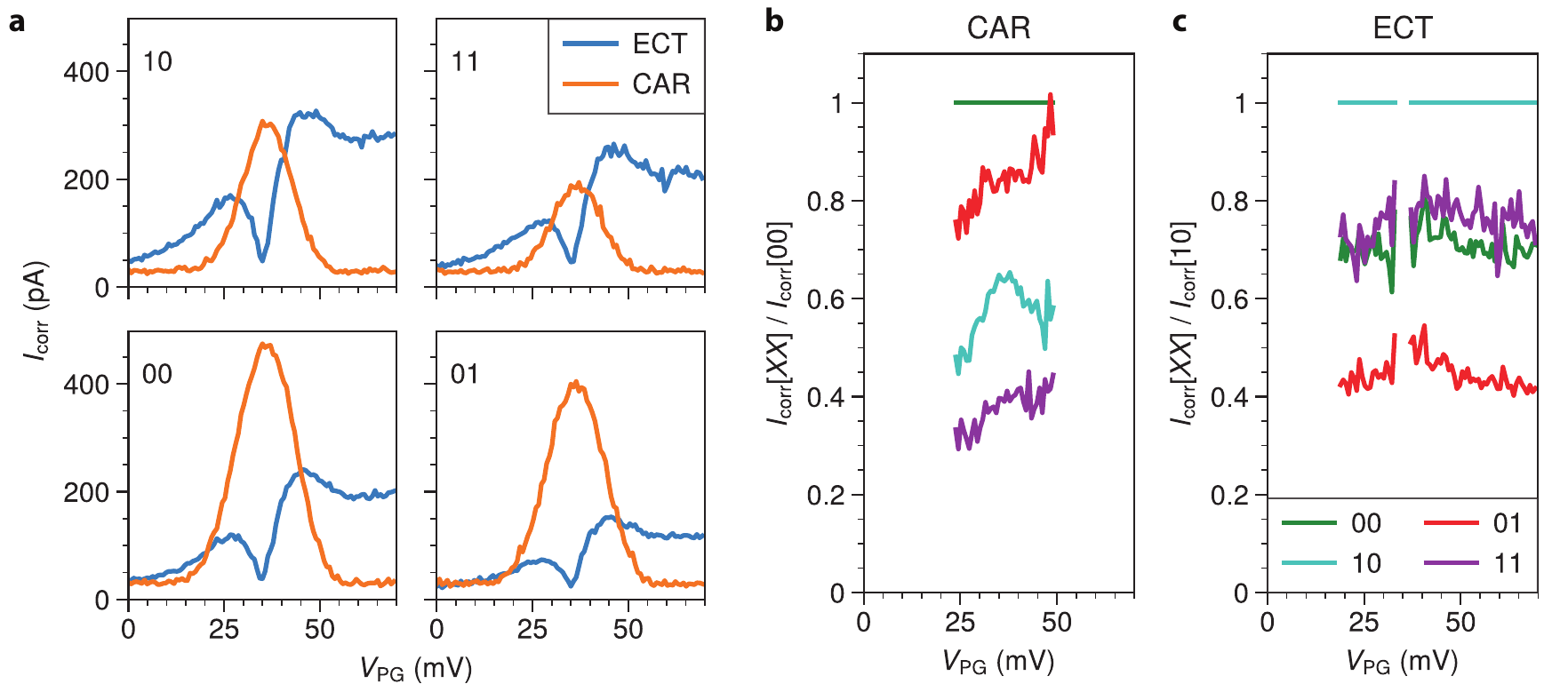}
    \caption{\textbf{Pauli spin blockade at $\vec{B} = 0$.}
    \textbf{a.} $\VPG$ dependence of CAR and ECT using four charge degeneracy points corresponding to one pair of spin-degenerate levels in each QD. Bias voltages are the same as described in the main text. Since there is no magnetic field, they are here denoted `00', `01', `10', `11', instead of `$\downarrow\downarrow$', `$\downarrow\uparrow$', `$\uparrow\downarrow$', `$\uparrow\uparrow$' to avoid confusion. The `00' data displayed in the bottom left plot is the same as Fig.~\ref{fig: single_ABS}f. All four charge degeneracy points show the same, characteristic curve shapes: single-peaked for CAR and double-peaked for ECT. For the bias polarities used here, Pauli spin blockade reduces the overall magnitude of CAR in the `11' charge degeneracy and that of ECT in the `01' charge degeneracy~\cite{Wang.2022}. 
    \textbf{b.} CAR magnitudes divided by that of the `00' charge degeneracy point. CAR-induced currents smaller than \SI{50}{pA} are excluded from the plot to avoid division by small numbers.
    \textbf{c.} ECT magnitudes divided by that of the `10' charge degeneracy point. ECT-induced currents smaller than \SI{50}{pA} are excluded from the plot to avoid division by small numbers. Panels~b and c show that the ratios of CAR and ECT magnitudes relative to the non-spin-blockaded process are roughly constant as a function of $\VPG$. Thus, although Pauli spin blockade is not part of the theoretical model, its effect is mainly an overall scaling of the CAR amplitude relative to ECT and does not alter the $\VPG$ dependence of them each.}
    \label{fig: zerofield_all}
\end{figure}

\begin{figure}[h]
    \centering
    \includegraphics[width = 1\textwidth]{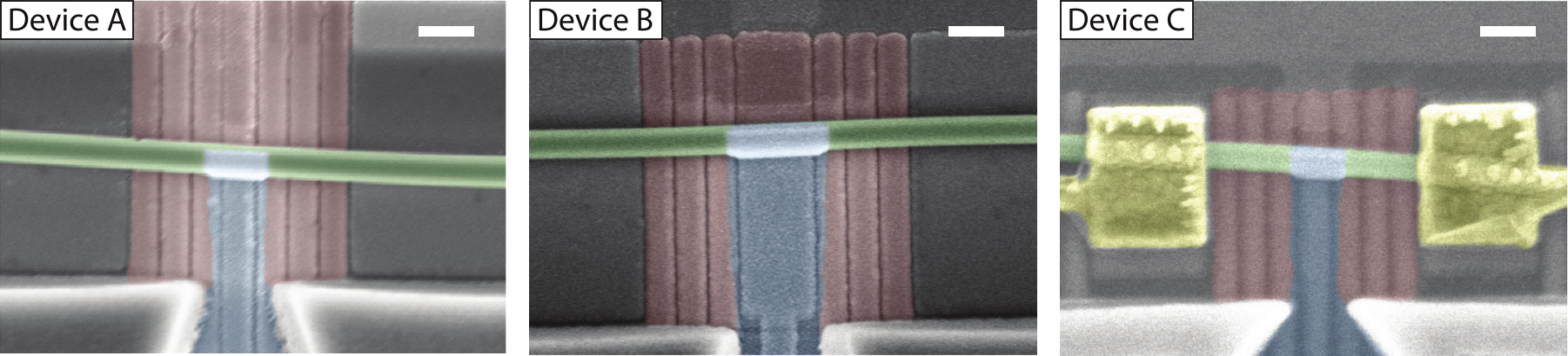}
    \caption{\textbf{False-colored SEM images of measured devices.} Green is nanowire, blue is Al (Al+Pt for device A), red are bottom gates and yellow are Au contacts. Devices A and B were imaged prior to Au contact deposition. Scale bars are $\SI{200}{nm}$.}
    \label{fig: SEMs}
\end{figure}

\begin{figure}[h]
    \centering
    \includegraphics[width = \textwidth]{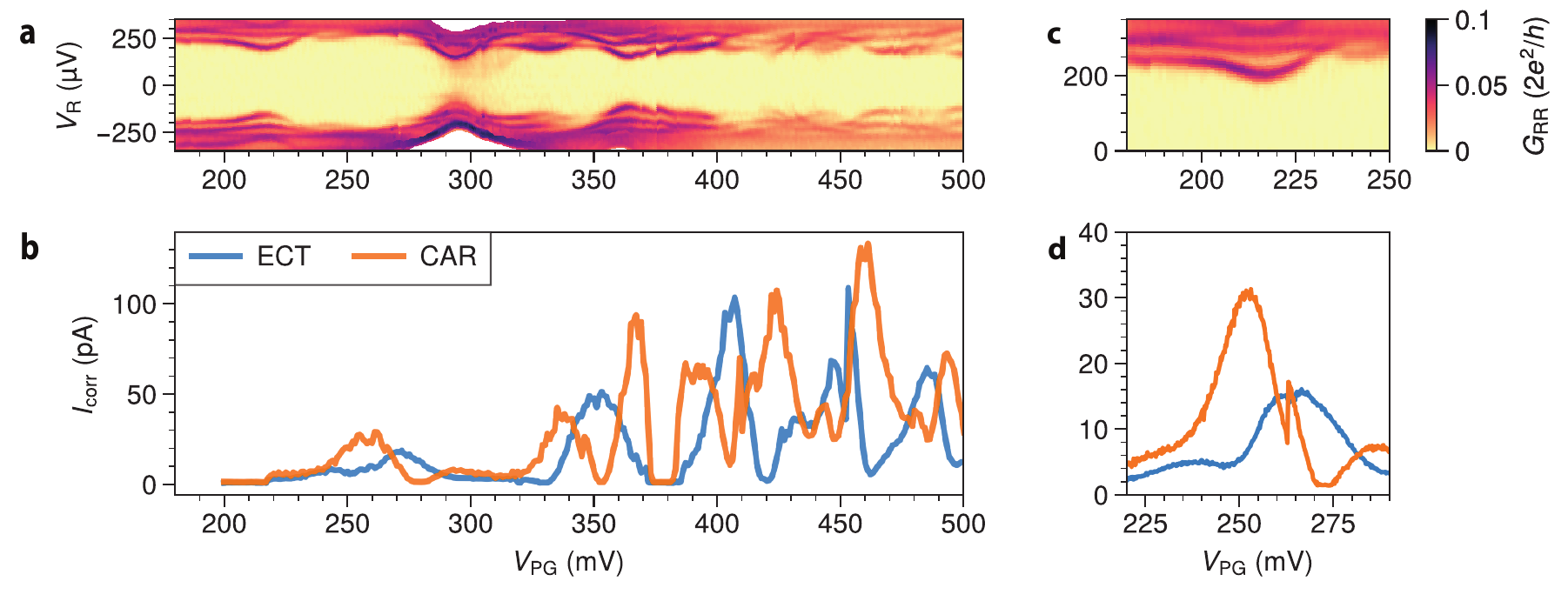}
    \caption{\textbf{Correlation between ABS and CAR/ECT processes in device B.} 
    \textbf{a.} Local spectroscopy of device B, measured with standard lock-in techniques. The voltage bias $\VL$ is corrected for a series resistance of \SI{105}{k\ohm} to take into account a fridge line resistance of \SI{2.9}{k\ohm} and a current meter resistance of \SI{102}{k\ohm}. Data is taken at $B = B_x = \SI{100}{mT}$. 
    \textbf{b.} CAR and ECT magnitudes as a function of $\VPG$. As opposed to what is presented for device A in Fig.~\ref{fig: Correlation}, the values of $V_\mathrm{LI}$ and $V_\mathrm{RI}$ are not the same for the measurements shown in panels a and b. This results in a shift of $\sim\SI{40}{mV}$ of the ABS positions with respect to $\VPG$, due to cross-coupling between neighboring gates. Data is taken at $B=0$. 
    \textbf{c.} Zoom in of panel~a around the first ABS. To compensate for the gate shift mentioned above, the plotted $\VPG$ ranges differ by $\SI{40}{mV}$ for easier comparison. 
    \textbf{d.} High resolution measurement of CAR and ECT magnitudes for the first ABS. The effect of a gate jump can be seen at $\VPG\approx 260$ mV. An ECT dip, signature of destructive interference, is visible at $\VPG = \SI{250}{mV}$, although it is less pronounced than what observed for device A in Fig.~\ref{fig: single_ABS}. A weaker interference might be due to the presence of multiple ABSs (a second ABSs is visible in panel~c for $\VPG\approx \SI{245}{mV}$). The smaller semiconducting level spacing likely results from device B being longer than device A: the hybrid sections are \SI{350}{nm} and \SI{200}{nm} long, respectively.}
    \label{fig: devB}
\end{figure}

\begin{figure}[h]
    \centering
    \includegraphics[width = 0.45\textwidth]{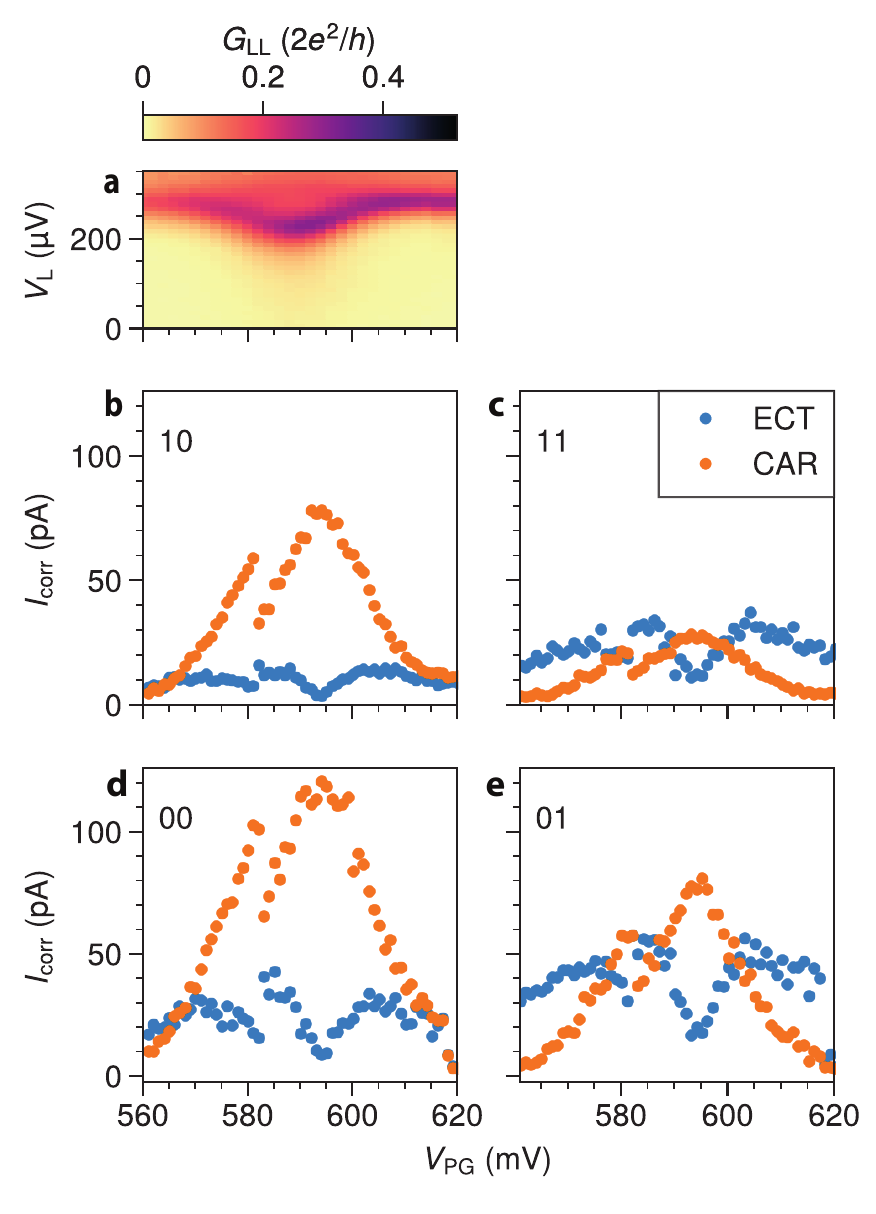}
    \caption{\textbf{ECT interference in another device (device C).} 
    \textbf{a.} Local spectroscopy of device C, measured as in Fig.~\ref{fig: single_ABS}. \textbf{b--e.} CAR and ECT magnitudes as a function of $\VPG$ for four charge degeneracy points. Destructive interference of ECT is visible at $\VPG = \SI{595}{mV}$. The discontinuity visible in all plots at $\VPG = \SI{582}{mV}$ is attributed to a gate jump.}
    \label{fig: devC}
\end{figure}

\begin{figure}[h]
    \centering
    \includegraphics[width = \textwidth]{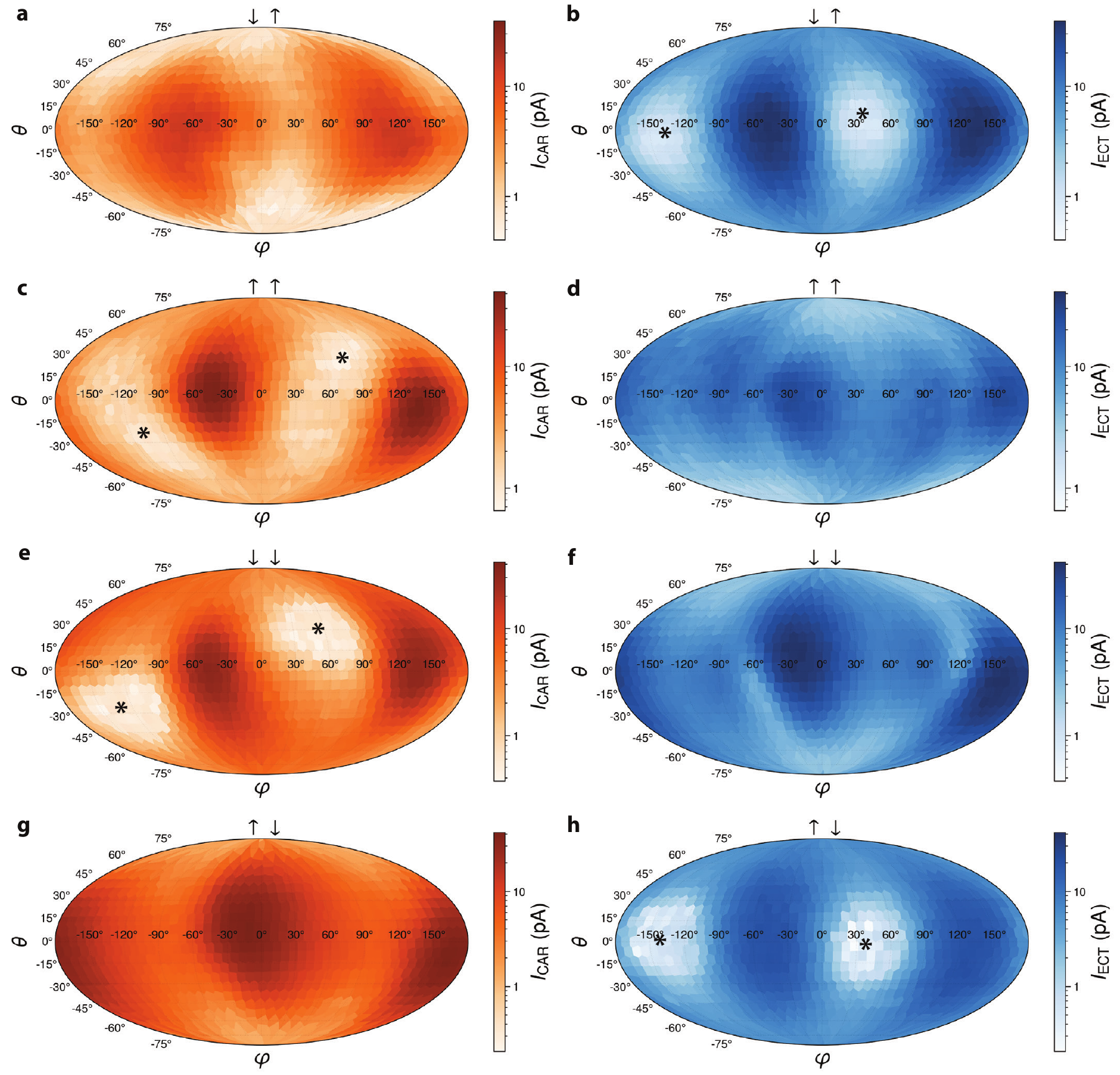}
    \caption{\textbf{CAR and ECT dependence on field direction for other spin selections in device B.} Spherical plots as in Fig.~\ref{fig: globes}a-b when the QDs select other spin configurations (the $\downarrow\uparrow$ configuration is reported here again in panels a and b as in the main text for easier comparison).
    When opposite spins are selected, ECT is suppressed along a single direction. While, when the QDs select $\uparrow\uparrow$ or $\downarrow\downarrow$ spins, it is the CAR-induced current to be suppressed along a single magnetic field direction. We interpret the suppression direction as the orientation of the spin orbit field $\vec{B}_\mathrm{SO}$ and highlight it with star marks. We remark that the suppression direction as well as the enhancement direction is slightly different among plots. The origin of this discrepancy is not yet fully understood. Following the discussion regarding ABS charge, we speculate that it could be caused by more than one ABS mediating ECT and CAR. 
    Concretely, the ABS most responsible for ECT could have a slightly different spin-orbit direction than the one mediating CAR.
    }
    \label{fig: globes-supp}
\end{figure}

\begin{figure}[h]
    \centering
    \includegraphics[width = 0.6\textwidth]{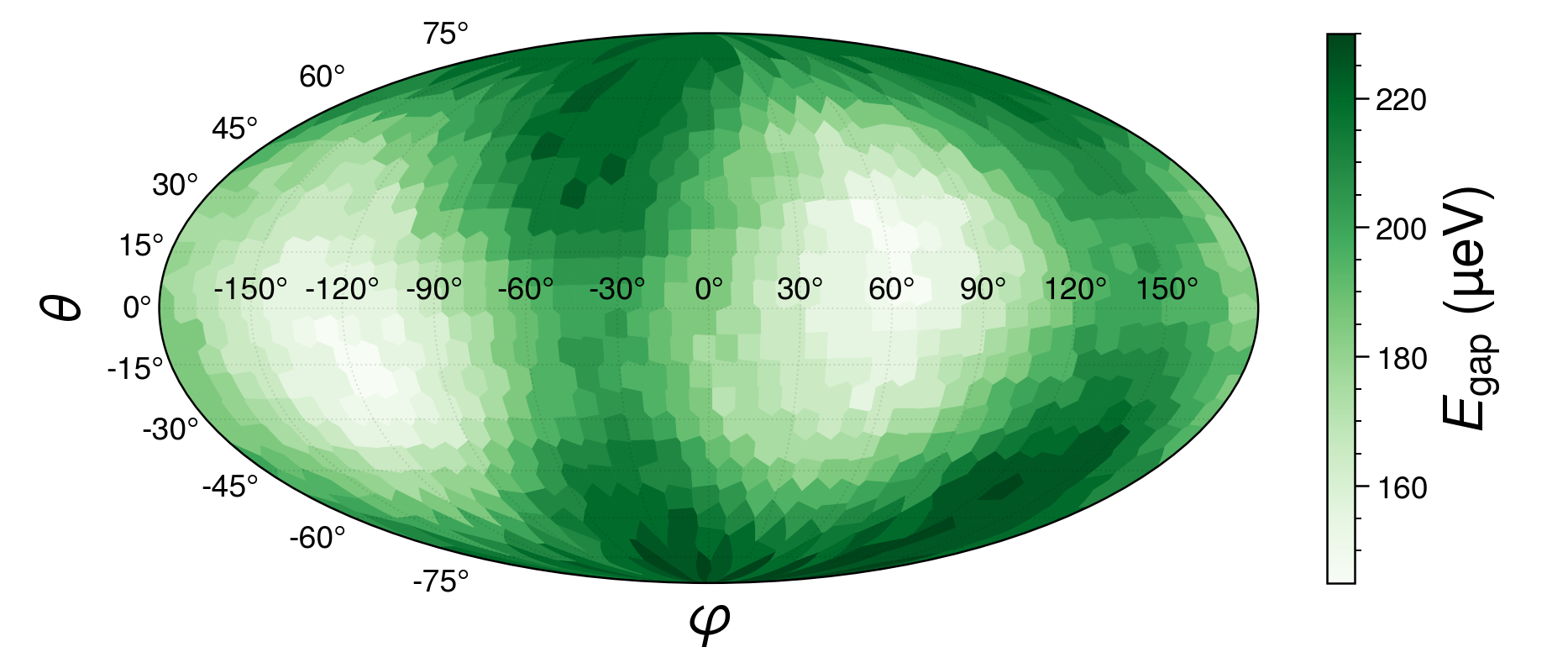}
    \caption{\textbf{Superconducting gap dependence on magnetic field direction.} 
    Local tunneling spectroscopy as in Fig.~\ref{fig: globes}c but at negative superconducting gate: $\VPG = \SI{-500}{mV}$. 
    Since no ABSs are present at this $\VPG$ value, this is a direct measurement of the hybrid superconducting gap as a function of magnetic field orientation. 
    $\abs{\vec{B}} = \SI{150}{mT}$. 
    Notably, the direction along which the gap is reduced the most is different from that along which the ABS of Fig.~\ref{fig: globes}c reaches its energy minimum. 
    The gap-suppression direction in this strongly metallized regime~\cite{Mazur.2022} is likely where orbital depairing in the Al film is the strongest, considering the size of $\abs{\vec{B}}$ and that it is the angle that maximizes the flux incident on the Al-covered facets.
    Superficially, previous work on hybrid nanowires has also observed maximal gap suppression along similar angles and interpreted it as the measured spin-orbit direction~\cite{Bommer.2019}.
    However, the analysis there relied on the superconducting film made of NbTiN experiencing almost no orbital depairing along all magnetic field directions.
    The same interpretation is not valid in the case of Al here and thus gap spectroscopy cannot be used to measure the effect of spin-orbit coupling.
    Therefore, using CAR and ECT to measure the spin-orbit direction is less prone to complications by orbital effects compared to gap-size spectroscopy.}
    \label{fig: ABS-globes-supp}
\end{figure}

\begin{figure}[h]
    \centering
    \includegraphics[width = 0.9\textwidth]{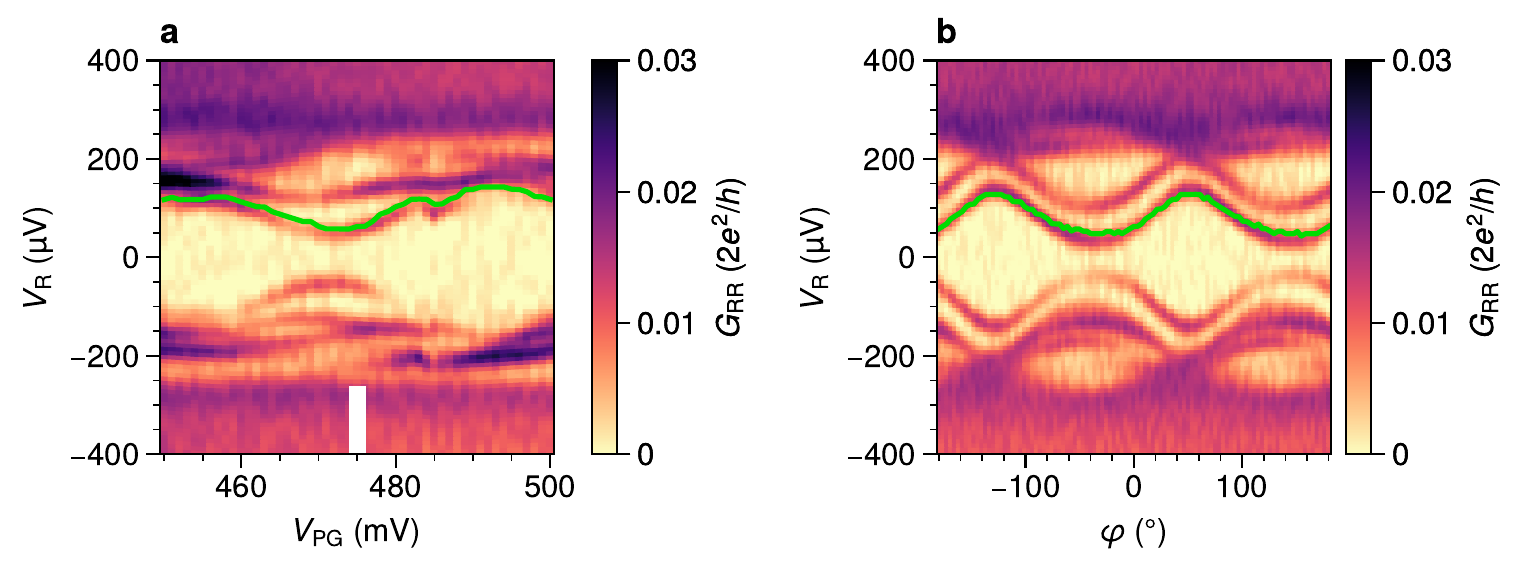}
    \caption{\textbf{Comparison between tunnel spectroscopy and extracted $E_\downarrow$.} 
    \textbf{a.} Local tunnel spectroscopy of device B as a function of $\VPG$ around the values used in Fig.~\ref{fig: globes}a--c, measured using a lock-in. $\abs{\vec{B}}=\SI{80}{mT}$, applied perpendicular to the nanowire ($\theta = 0^\circ$) and with $\varphi= -180^\circ$. Green: extracted $E_\downarrow$ values using the same method as in Fig.~\ref{fig: globes}c. As in that panel, the calculation is done using measured currents, in this case simultaneously acquired as the lock-in conductance. White bar marks the $\VPG$ value at which panel~b data is taken.
    \textbf{b.} Idem, as a function of field angle $\varphi$. 
    }
    \label{fig: ED_gapsize_spec}
\end{figure}


\end{document}